\documentclass[sigconf]{acmart}

\usepackage{xifthen}
\usepackage{multirow}
\usepackage{verbatim}
\usepackage{amsmath}

\usepackage{subcaption}
\PassOptionsToPackage{hyphens}{url}
\usepackage[hyphens]{url}
\usepackage{hyperref}
\usepackage{booktabs}
\usepackage{graphicx}
\usepackage{caption}
\usepackage{pdfpages}
\usepackage{setspace}
\usepackage{subfiles}
\usepackage[font=small,labelfont=bf]{caption}

\date{} 
\setcopyright{none}
\renewcommand\footnotetextcopyrightpermission[1]{} % removes footnote with conference information in first column
\begin{document}

% Copyright
%\setcopyright{acmcopyright}
%\setcopyright{acmlicensed}
%\setcopyright{rightsretained}
%\setcopyright{usgov}
%\setcopyright{usgovmixed}
%\setcopyright{cagov}
%\setcopyright{cagovmixed}

% DOI
%\doi{10.475/123_4}

% ISBN
%\isbn{123-4567-24-567/08/06}

%Conference
%\conferenceinfo{PLDI '13}{June 16--19, 2013, Seattle, WA, USA}

%\acmPrice{\$15.00}

%
% --- Author Metadata here ---
%\conferenceinfo{WOODSTOCK}{'97 El Paso, Texas USA}
%\CopyrightYear{2007} % Allows default copyright year (20XX) to be over-ridden - IF NEED BE.
%\crdata{0-12345-67-8/90/01}  % Allows default copyright data (0-89791-88-6/97/05) to be over-ridden - IF NEED BE.
% --- End of Author Metadata ---

\title{Unimem: Runtime Data Management on Non-Volatile Memory-based Heterogeneous Main Memory} %for High Performance Computing}

\author{Kai Wu}
\affiliation{%
  \institution{University of California, Merced}
}
\email{kwu42@ucmerced.edu}

\author{Yingchao Huang}
\affiliation{%
  \institution{University of California, Merced}
}
\email{yhuang46@ucmerced.edu}

\author{Dong Li}
\affiliation{%
  \institution{University of California, Merced}
}
\email{dli35@ucmerced.edu}

\begin{abstract}  %150 words limitation for SC'17
Non-volatile memory (NVM) provides a scalable and power-efficient solution to replace DRAM as main memory. However, because of relatively high latency and low bandwidth of NVM, NVM is often paired with DRAM to build a heterogeneous memory system (HMS). As a result, data objects of the application must be carefully placed to NVM and DRAM for best performance. In this paper, we introduce a lightweight runtime solution that automatically and transparently manage data placement on HMS without the requirement of hardware modifications and disruptive change to applications. Leveraging online profiling and performance models, the runtime characterizes memory access patterns associated with data objects, and minimizes unnecessary data movement. Our runtime solution effectively bridges the performance gap between NVM and DRAM. We demonstrate that using NVM to replace the majority of DRAM can be a feasible solution for future HPC systems with the assistance of a software-based data management.

\end{abstract}

\maketitle

\section{Introduction}
\label{sec:intro}
%The emergence of NVM. \\
Non-volatile memory (NVM), such as phase change memory (PCM) and resistive random-access memory (ReRAM), is a promising technique to build future high performance computing (HPC) systems.
The popularity of many-core platforms in HPC %(such as Intel Xeon Phi on Aurora and Cori HPC)
and large data sets in scientific simulations drive the fast development of NVM,
because NVM can provide a scalable and power-efficient solution as main memory,
alternative to DRAM.
Such solution is based on the attractive characteristics of NVM, such as higher density and near-zero static power consumption.

However, comparing with DRAM, NVM as main memory can be challenging.
The promising NVM solutions (e.g., PCM and ReRAM), 
although providing larger capacity at the similar or lower cost than DRAM,
can have higher latency and lower bandwidth (see Table~\ref{tab:nvm_features}).
Such NVM features can introduce a big performance gap between emerging NVM-based and traditional DRAM-based systems for HPC applications. 
Our initial performance evaluation with HPC workloads (Section~\ref{sec:bg}) shows that there is 1.09x-8.4x slowdown on NVM-based systems, depending on bandwidth and latency features of NVM.
Because of the limitation of NVM, NVM is often paired with a small fraction of DRAM to form a heterogeneous memory system (HMS)~\cite{eurosys16:dulloor, nas16:giardino, asplos16:lin, ismm16:shen, gpu_pcm_pact13, hpdc16:wu}.
%Leveraging such HMS, 
By selectively placing frequently accessed data in the small amount of DRAM available in HMS, we are able to exploit the cost and scaling benefits of NVM while minimizing the limitation of NVM with DRAM.

\begin{comment}
The emergence of the NVM-based HMS breaks the uniformity in the traditional memory system:
%the previous assumption that all memory devices have the similar latency and bandwidth is not true.
%and introduces a gap between application and memory:
As a result, the existing system must be evolved to handle such memory heterogeneity.
Data objects of the application must be carefully placed to NVM and DRAM
for best performance.
%Mapping data objects of the application to memory is not straightforward as in the traditional system. Instead, we must carefully place those data objects on HMS for best performance.
%In fact, such breaking in memory uniformity is not new: we already see that in NUMA systems. But the emergence of NVM exacerbates this problem.
\end{comment}

To manage data placement on HMS for HPC, we have several goals.
%for the system design. 
First, we want to avoid disruptive changes to hardware. The existing hardware-based solutions to manage data placement on HMS~\cite{qureshi_micro09, ibm_isca09, gpu_pcm_pact13, row_buffer_pcm_iccd12} may be difficult to be embraced by
HPC data centers, because of the concerns on hardware cost.
Second, we want to minimize changes to applications and system software.
HPC legacy applications should be easily ported to NVM-based HMS
with few programming efforts.
Third, managing data placement should be as transparent as possible.
We want to enable automatic data placement, and relieve users from
managing data placement details.

%we have hardware- and software-based solutions. 
%The hardware-based solutions introduce disruptive changes to existing hardware and system software. Such changes may result in larger investment to update existing HPC data centers On the other hand, the software-based solution . 
%We have see the success story of using algorithm knowledge~\cite{} and classifying memory access patterns for perf, energy, and programming models for data persistence

In this paper, we introduce a software-based solution to decide
and place data objects on NVM-based HMS. Using a software-based solution
to meet the above goals must address the following research challenges. 
First, how to capture and characterize memory access patterns associated with data objects? This question is important for making data placement decisions.
As we show in Section~\ref{sec:bg},
after we move some data object from %bandwidth-limited 
NVM with less memory bandwidth 
to DRAM, there is a big performance improvement.
However, we do not have such performance improvement after moving
this data object from NVM %latency-limited
with longer access latency to DRAM.
We claim such data object is sensitive to memory bandwidth.
Similarly, we find some data object which is only sensitive to memory latency, or sensitive
to both bandwidth and latency. 
%some data objects have performance of
%their data placement sensitive to memory bandwidth, while some data objects
%have performance of their data placement sensitive to memory latency. 
Characterizing data objects 
based on their sensitivity to bandwidth or latency
is critical to model and predict performance benefit of data placement. 
%Furthermore, to make our software-based solution feasible in HPC, we do not want to modify existing hardware and application, or employ offline analysis~\cite{ismm16:shen, eurosys16:dulloor} to characterize data objects. 

Second, how to strike a balance between different requirements
on the frequency of data movement (i.e., the implementation of data placement)?
On one hand, we want data movement to be frequent, %enough, 
such that data placement is adaptive to variation of memory access patterns across execution phases.
On the other hand, we want to minimize data movement to avoid performance loss.

Third, how to minimize the impact of data movement on application performance? Data movement is known to be expensive
in terms of performance and energy cost. 
Hiding data movement cost and achieving high performance is
a key to be successful in the HPC domain.

In this paper, we introduce a runtime system (named ``Unimem'') that automatically and transparently decides and implements data placement.
This runtime meets the above goals and addresses the above three challenges.
In particular, we employ online profiling based on performance counters
to capture memory access patterns for execution phases, based on which we characterize the sensitivity of data objects in each phase to memory bandwidth and latency.
This addresses the first challenge.
We further introduce lightweight performance models, based on which we predict 
performance benefit and cost if moving data objects between NVM and DRAM.
Given the performance benefit and cost of data movement, we formulate the problem of deciding optimal data placement as a a knapsack problem.
Based on the performance models and formulation, we avoid unnecessary data movement
while maximizing the benefits of data movement.
This addresses the second challenge.

To avoid the impact of data movement on application performance, we introduce a proactive data movement mechanism. Given an execution phase and a data movement plan for the phase,
this mechanism uses a helper thread to trigger data movement before the phase.
The helper thread runs in parallel with the application, overlapping data movement with application execution. This proactive data movement mechanism takes data movement overhead off the critical path, which addresses the third challenge.
To further improve performance, we introduce a series of
techniques, including (1) optimizing initial data placement to reduce data movement cost at runtime, (2) %dynamic phase delineation to 
exploring the tradeoff between phase local search and cross-phase global search for optimal data placement, and (3) decomposing large data objects to enable fine-grained data movement.
All together, those techniques in combination with our performance models
greatly narrow the performance gap between NVM and DRAM:
%The performance difference between DRAM-only and HMS with Unimem
%is 7\% on average and 13\% at most for representative HPC workloads.

In summary, we make the following contributions.
%\linebreak
%\vspace{-4pt}
\begin{itemize}
    \item We study performance of HPC workloads with large data sets on multiple nodes with various NVM bandwidth and latency, which is unprecedented. 
    Our study reveals a big performance gap between NVM-based and DRAM-based main memories. 
    We demonstrate the feasibility of using a runtime-based solution 
    %to direct data placement on HMS
    to narrow such gap for HPC.
    \item We introduce a lightweight runtime system to manage data placement without
    %the requirement of 
    hardware modifications and disruptive changes to applications and system software.
    \item We evaluate Unimem with six representative HPC workloads and one production code (Nek5000). The performance difference between DRAM-only and HMS with Unimem is only 6.2\% on average and 16\% at most. We successfully narrow the performance gap and demonstrate better performance than a state-of-the-art software-based solution.
\end{itemize}

\begin{comment}
Different GPU hetero, there is no programm challening.
But performance challenges
The challenges to use the heterogeneous memory.
(1) bandwidth vs. latency; (2) runtime overhead;
(3) must be adaptive to the application phase changes
These execution phases are sensitive to NVM higher latency or lower bandwidth or both.
Determine the placement is challenging. Runtime overhead
\\

This study has clearly demonstrated that runtime data management

The runtime provides a transparent runtime adaptation of HPC codes,
and requires only a trivial code instrumentation step.
\end{comment}

\section{Background}
\label{sec:bg}
%The basic assumption on the memory architecture. (DRAM and NVM side-by-side)
%To overcome the limitation of NVM-only system, future systems
%are likely to build HMS by coupling NVM with a smaller amount of DRAM.
In HMS, we assume that DRAM shares the same physical address space as NVM (but with different addresses) and DRAM memory allocation can be managed at the user level. This assumption has been widely used in the existing work~\cite{eurosys16:dulloor, nas16:giardino, asplos16:lin, ismm16:shen,  hpdc16:wu}. %We assume such system in this paper.

\subsection{Definitions and Basic Assumptions}
%What are phases? Definition of phases (check ICS'09 (LLNL) power "Adagio: DVS practical") \\
%How to handle non-blocking MPI call as phases? How to handle MPI_Wait\\
%Draw a figure to explain the idea of "phases" if necessary;
%We target a SPMD programming model on distributed memory systems using 
%message passing, particularly MPI, for any communication between processes.
%For a parallel application based on such programming model,
We target on the MPI programming model.
For a parallel application based on MPI, we decompose the application into phases.
A phase can be a computation phase delineated by MPI operations;
A phase can also be an MPI communication phase doing collective operations, point-to-point communication operations, 
or synchronization. For a non-blocking communication (e.g., {\fontfamily{qcr}\selectfont MPI\_Isend}), the MPI communication call 
is not a phase. Instead, it is merged into 
the immediately following phase. 
The communication completion operation (e.g., {\fontfamily{qcr}\selectfont MPI\_Wait}) is a communication phase. 

\begin{figure}
\centering
\includegraphics[width=0.2\textwidth, height=0.25\textheight]{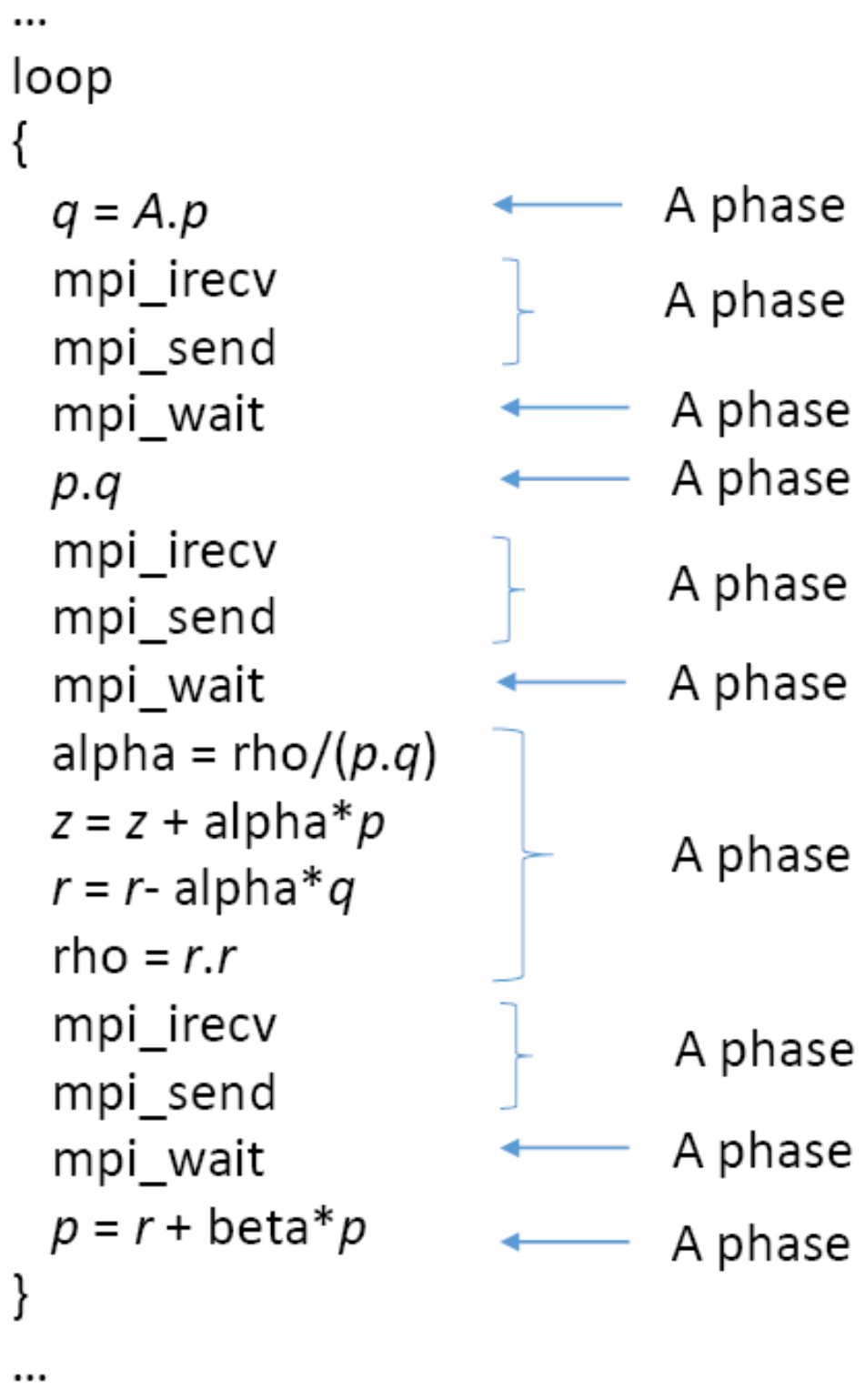}
\vspace{-5pt}
\caption{A conceptual description for an MPI-based program (CG) decomposed into phases. $A$ is a 2D matrix, and $q, p, z$, and $r$ are vectors.}
\label{fig:phases_gen_desc}
\vspace{-15pt}
\end{figure}

Furthermore, we target on parallel applications from the HPC domain with
an iterative structure. In those applications, each program phase is executed many times.
Such parallel applications are very common. 
As an example, Figure~\ref{fig:phases_gen_desc} depicts a typical iterative structure from CG (an NAS parallel benchmark~\cite{nas}), which dominates the execution time of CG.

We claim a data object is~\textit{bandwidth sensitive}, if there is a big performance difference between placing it in %bandwidth-limited NVM and DRAM.
NVM with less memory bandwidth and DRAM.
We claim a data object is~\textit{latency sensitive}, if there is
a big performance difference between placing it in %latency-limited NVM and DRAM.
NVM with longer memory access latency and DRAM.

%\subsection{Performance Characteristics of Numerical Algorithms on NVM}
\subsection{Preliminary Performance Evaluation with NVM-Based Main Memory}
NVM has relatively long access latency and low memory bandwidth.
Table~\ref{tab:nvm_features} shows NVM performance characteristics.
The table is based on~\cite{NVMDB} that gathered a comprehensive survey of 340 non-volatile memory technology papers published between 2000 and 2014 in relevant conferences.
Based on such performance characteristics, we perform preliminary performance study
to quantify the impact of NVM on HPC application performance.
%Also, we want to quantify the performance gap between NVM-only and DRAM-only main memories.
%Such study can quantify the application performance gap between NVM- and DRAM-based main memory.

\begin{comment}
\begin{table*}
        \tiny
        \begin{center}
            \caption{NVM performance characteristics, and comparison between NVM techniques and DRAM~\cite{NVMDB}}
            \vspace{-10pt}
       \label{tab:nvm_features}
       \begin{tabular}{|p{2.4cm}|p{1.8cm}|p{1.8cm}|p{2cm}|p{2.5cm}|p{2cm}|p{2cm}|}
       \hline
              & \textbf{Read time (ns)}  & \textbf{Write time (ns)} &   \textbf{32B random read BW (MB/s)} & \textbf{32B random write BW (MB/s)} & \textbf{Seq read BW (MB/s)} &   \textbf{Seq write BW (MB/s)}                  \\ \hline \hline
                                              DRAM       & 10    & 10 & 1,000 & 900 & 11,000 & 11,000        \\\hline
                                              STT-RAM (ITRS'13)      & 60 & 80 & 800 & 600 & 11,000 & 11,000                                 \\ \hline
                                              PCRAM        & 20-200 & 80-10,000 & 200-800 & 100-800 & 11,000 & 2,000-11,000                            \\ \hline
                                            ReRAM       & 10-1,000 & 10-10,000 & 20-100 & 1-8 & 10,000-11,000 & 900-8,000               \\ \hline
      \end{tabular}
      \end{center}
\end{table*}
\end{comment}

\begin{table}
        \tiny
        \begin{center}
            \caption{NVM performance characteristics, and comparison between NVM techniques and DRAM~\cite{NVMDB}}
            \vspace{-10pt}
       \label{tab:nvm_features}
       \begin{tabular}{|p{0.8cm}|p{1.2cm}|p{1.2cm}|p{1.5cm}|p{1.5cm}|}
       \hline
              & \textbf{Read time}  & \textbf{Write time} &   \textbf{Random read BW} & \textbf{Random write BW}    \\ \hline \hline
                                              DRAM       & 10 ns    & 10 ns & 1,000 MB/s & 900 MB/s        \\\hline
                                              STT-RAM (ITRS'13)      & 60 ns & 80 ns & 800 MB/s & 600 MB/s                                  \\ \hline
                                              PCRAM        & 20-200 ns & 80-10,000 ns & 200-800 MB/s & 100-800 MB/s                            \\ \hline
                                            ReRAM       & 10-1,000 ns & 10-10,000 ns & 20-100 MB/s & 1-8 MB/s            \\ \hline
      \end{tabular}
      \end{center}
      \vspace{-10pt}
\end{table}

\begin{figure}[!t]
    \centering
    \includegraphics[width=0.48\textwidth, height=0.13\textheight]{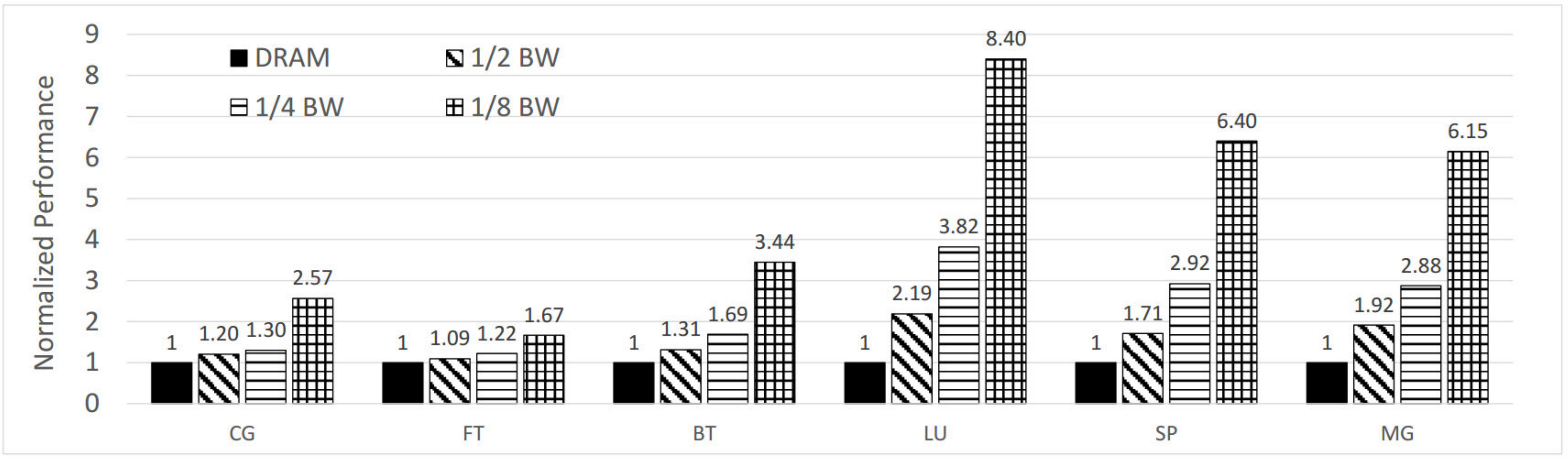}
    \caption{The benchmark performance (execution time) on NVM-based main memory (NVM-only) with various bandwidth. The performance is normalized to that of DRAM-only systems.} 
    \label{fig:bg_bw_sensitivity}
\end{figure}

\begin{figure}[!t]
    \centering
    \includegraphics[width=0.48\textwidth, height=0.13\textheight]{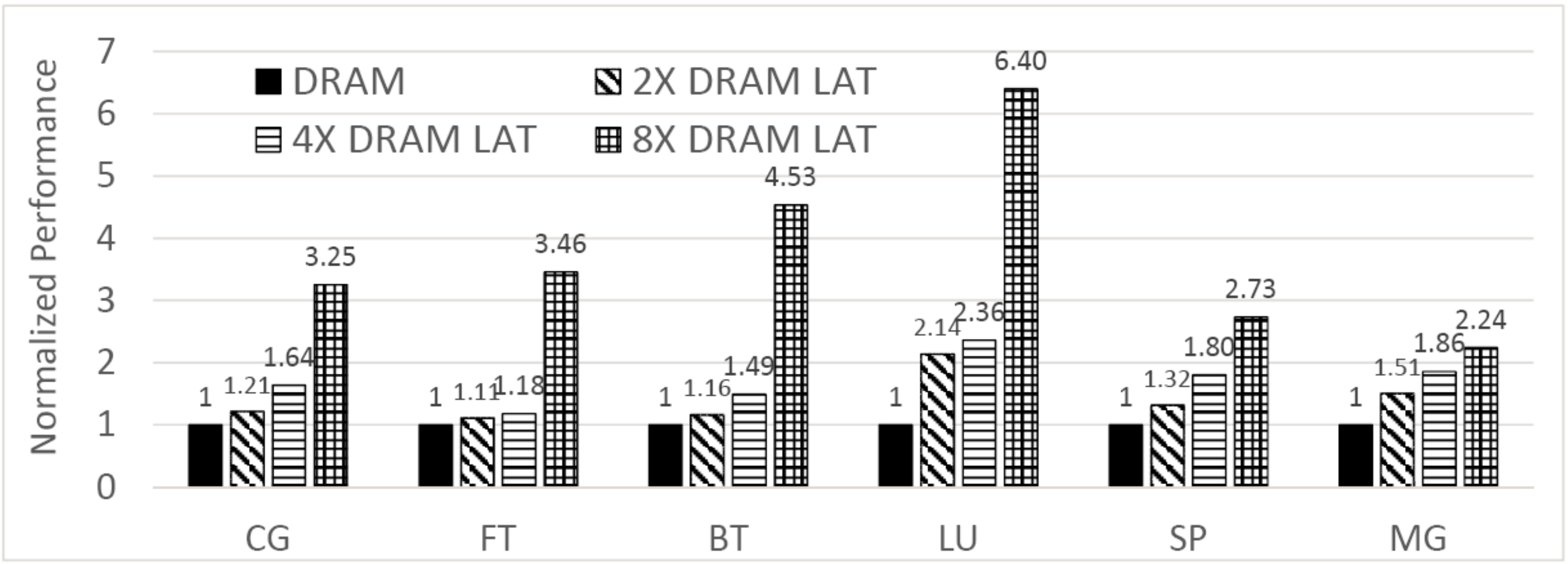}
	    \caption{The benchmark performance (execution time) on NVM-based main memory (NVM-only) with various latency. The performance is normalized to that of DRAM-only systems.} 
    \label{fig:bg_lat_sensitivity}
\end{figure}

We use Quartz, a DRAM-based, lightweight performance emulator for NVM~\cite{middleware15:volos}. %for our study. 
The existing work uses cycle-accurate simulation to study NVM performance~\cite{nvm_ipdps12, hpdc16:wu}.
However, the long simulation time makes impossible simulate~\textit{HPC applications} with large data sets on multiple nodes.
The performance of HPC workloads on NVM is always mysterious.
Using Quartz, we can study performance (execution time) of HPC workloads with much shorter time. 
We deploy our tests on four nodes in Platform A (the configurations of those nodes and Platform A are summarized in Section~\ref{sec:eva}).
We change the emulated NVM bandwidth and latency, and run a set of NAS parallel benchmarks. We use Class D as input and run 16 MPI processes (4 MPI processes per node). For the benchmark FT, we use CLASS C as input because of the long execution time with Class D.  Figures~\ref{fig:bg_bw_sensitivity} and~\ref{fig:bg_lat_sensitivity} show the emulation results.
%In the rest of the paper, we call a data object that has the performance of its data placement sensitive to memory bandwidth (or latency) as \textit{bandwidth sensitive} (\textit{latency sensitive}) data object. 

%Evaluation 1: changes latency and bandwidth and study the performance sensitivity of benchmarks and applications \\
%Evaluation 2: the performance sensitivity of data objects with different data placement 

\begin{figure}[!t]
    \centering
    \includegraphics[width=0.48\textwidth, height=0.16\textheight]{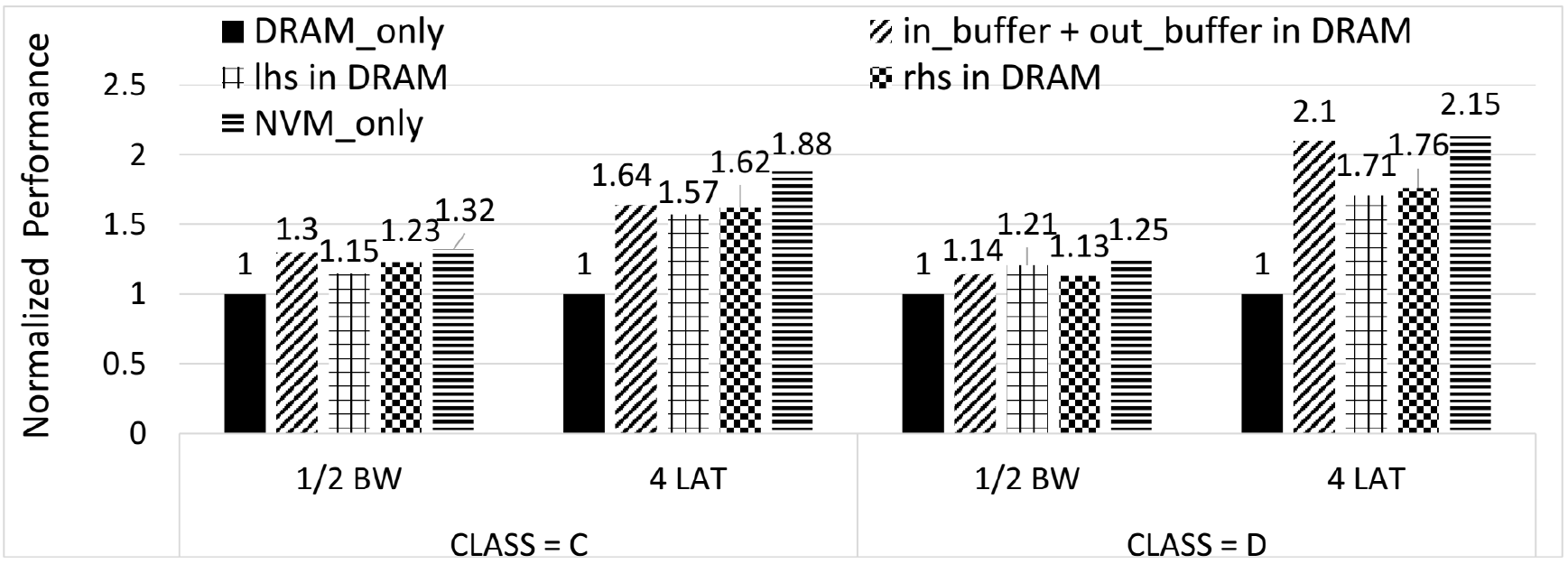}
    \caption{The impact of data placement on performance (execution time) of NVM-based main memory. The performance is normalized to DRAM-only systems. The legend entries ``in\_buffer+out\_buffer'', ``lhs'', and ``rhs'' are the data objects placed in DRAM in the DRAM+NVM system. The $x$ axis shows the configuration of NVM (4x DRAM latency or 1/2 DRAM bandwidth).}
    \label{fig:bg_data_placement_impact}
    \vspace{-15pt}
\end{figure}

\textbf{Observation 1:}
We find a big performance gap between DRAM-only and NVM-only systems.
%for HPC workloads in the two figures. 
This observation is contrary to an existing conclusion (i.e., no big gap) for HPC workloads 
based on a single node simulation~\cite{nvm_ipdps12}.
Furthermore, HPC application performance (execution time) is sensitive to different NVM technologies with various bandwidth and latency. With memory bandwidth reduced by only 1/2 or latency increased by only 2x in NVM, some benchmarks already show big slowdown. For example, LU has 2.19x and 2.14x slowdown with NVM configured with 1/2 DRAM bandwidth (Figure~\ref{fig:bg_bw_sensitivity}) and 2x DRAM latency (Figure~\ref{fig:bg_lat_sensitivity}) respectively.
%the benchmarks have 10\%-218\% performance loss; with 1/8 of DRAM bandwidth, the benchmarks have 68\%-1122\% performance loss. With 4x of DRAM latency, the benchmarks have 2.1\%-56.2\% performance loss.  

%those benchmarks are particularly sensitive to memory bandwidth reduction. 

We further study whether data placement in HMS can bridge the performance gap between DRAM-based and NVM-based systems. We choose SP benchmark and focus on four critical data objects of SP (the arrays $lhs$, $rhs$, $in\_buffer$ and $out\_buffer$). We use two configurations for NVM, one with 1/2 DRAM bandwidth and the other with 4x DRAM latency.
%For each NVM configuration, we do four tests, each of which places one of the four
%data objects in DRAM while placing the other data objects in NVM. 
For each data object with an NVM configuration (either 1/2 DRAM bandwidth or 4x DRAM latency),
we do three tests. In the first test we use DRAM-only system. In the second test we use
a DRAM+NVM system. For this test, a target data object is placed in DRAM (see the legend entries in Figure~\ref{fig:bg_data_placement_impact}), while the rest of
data objects are placed in NVM. In the third test we use an NVM-only system.
In each test, we use 4 nodes with one MPI task per node, and use CLASS C and CLASS D as input.
Figure~\ref{fig:bg_data_placement_impact} shows the results. 
The results are normalized to the performance of DRAM-only.
%With the NVM configurations of 1/2 DRAM bandwidth and 4x DRAM latency, there are 25\% and 92\% performance loss respectively.

\textbf{Observation 2:}
A good data placement can effectively bridge the performance gap.
For example, with the data object $lhs$ placed in DRAM, we bridge the performance gap between DRAM and NVM (using the configuration of 4x DRAM latency and CLASS C) by 31\% (see Figure~\ref{fig:bg_data_placement_impact}).

\textbf{Observation 3:}
%To improvement performance, different data objects have different preference to data placement.
Different data objects manifest different sensitivity to
limited NVM bandwidth and latency, shown in Figure~\ref{fig:bg_data_placement_impact}. 
For example, for the data objects $in\_buffer$ and $out\_buffer$ (CLASS D), 
there is no big performance difference (2.1 vs. 2.15) between placing them in DRAM and placing them in NVM configured with 4X DRAM latency; 
%after moved from NVM (configured with 4X DRAM latency) to DRAM, does not result in performance improvement.
However, 
%if $rhs$ is placed from NVM (configured with 1/2 DRAM bandwidth) to DRAM,
%we bridge the performance gap between NVM-only and DRAM-only by 48\%.
there is a big performance difference (1.14 vs. 1.25) between placing them in DRAM and placing them in NVM configured with 1/2 DRAM bandwidth (CLASS D). 
This indicates that the two data objects are sensitive to memory bandwidth but not memory latency. 
$lhs$ (CLASS D) tells us a different story:
it is sensitive to latency (1.71 vs. 2.15), but not bandwidth (1.21 vs. 1.25). 
Also, $rhs$ is sensitive to both latency and bandwidth.
%due to their bandwidth and latency requirement.

Different data objects have different memory access patterns which manifest
different sensitivity to bandwidth and latency.
A data object with a memory access pattern of bad data locality and massive, concurrent memory accesses (e.g., streaming pattern) is sensitive
to memory bandwidth, while a data object with a memory access pattern of bad data locality and dependent memory accesses (e.g., pointer-chasing) is sensitive to memory latency.

\begin{comment}
For example, the data object $p$ is not sensitive to bandwidth and latency limitation of NVM: its data placement in NVM and DRAM does not matter a lot to performance ($\le$ 1.2\% performance difference). 
On the other hand, the data object $a$ is sensitive to bandwidth but not latency (Class D): its data placement in DRAM improves performance by 11\%.
Also, the data object $a$ is sensitive to both bandwidth and latency when using
Class C: its data placement in DRAM improves performance by 13\% (1/2 DRAM bandwidth) and 27\% (4x DRAM latency) respectively.

Further study reveals that the memory accesses to $a$ are dominated by
a streaming memory access pattern and such pattern does not have data dependency between memory accesses. A large memory bandwidth is beneficial for such data object with concurrent memory accesses and few data locality.
This above fact is especially pronounced when we use a larger input problem (Class D).
The memory accesses to $p$ are dominated by indirect data references ($p[colidx[i]]$),
and manifest a random memory access pattern with data reuse. 
\textbf{The size of $p$ is also relatively small (comparing with the last level cache size). As a result, a large body of memory references to $p$ happen in the cache hierarchy. Hence, $p$ is not sensitive to the data placement.}
\end{comment}

Our preliminary performance study highlights the importance of capturing memory
access patterns of data objects. It also shows us that it is possible to bridge the performance gap between NVM and DRAM by appropriately directing data placement on HMS.

\vspace{-10pt}

\section{Design and Implementation} 
\label{sec:sys_des}
%Our preliminary performance study reveals that 
%placing data objects appropriately on NVM and DRAM
%is very beneficial to bridge the performance gap between NVM and DRAM.
Motivated by the preliminary performance study, we introduce a runtime system (named ``Unimem'') targeting on directing data placement on HMS for HPC applications.

%\subsection{Unimem Overview}
%\label{sec:overview}
\begin{comment}
To make Unimem a runtime for HPC applications, we have several design goals.
First, Unimem must be lightweight. This means that 
changing data placement (i.e., data movement) should have minimum impact on application performance.
This also means that the process to determine data placement
must be lightweight.
%collecting application information and making data placement decision  before data movement must be lightweight.
Second, enforcing data placement should be transparent to applications.
We want to enable automatic data movement, and relieve users from
handling data movement details.
Third, we want to avoid any disruptive change to applications,
such that legacy applications can be easily ported to NVM-based HMS
with few programming efforts. 
%Third, Unimem should be portable across common HPC platforms and scalable.
\end{comment}

Unimem directs data placement for data objects (e.g., multi-dimensional arrays). The data objects must be %can be the critical data structures 
allocated using certain Unimem APIs by the programmer.
%can be either specified by the programmer using certain Unimem API.  %or allocated through regular heap memory allocation.
We call those data objects, the \textit{target data objects}, in the rest of the paper.
Unimem is phase based. It decides and changes data placement for target data objects for each phase based on runtime profiling and lightweight performance models.

In particular, Unimem profiles memory references to target data objects
with a few invocations of each phase. 
Then Unimem uses performance models to predict performance benefit and cost of data placement,
and formulates the problem of deciding optimal data placement as a knapsack problem. 
The results of the performance models and formulation direct data placement for each phase in the rest of the application execution.
%To improve runtime performance, we introduce a series of optimization techniques.
%such as proactive data movement, optimizing initial data movement, and decomposing large data objects.
We describe the design and implementation details in this section.

\subsection{Design}
\label{sec:design}
Unimem includes three %fours 
steps in its workflow: phase profiling, performance modeling, and data placement decision and enforcement.
%phase profiling, performance modeling, data placement decision, and data placement enforcement.
The phase profiling happens in the first iteration of the main computation loop of the application.
At the end of the first iteration, we build performance models and
make data placement decision. After the first iteration, we enforce
the data placement decision for each phase.
We describe the three %four 
steps in details as follows.
%(1) Profiling --- hw counter 
%(2) data placement decision  --- bandwidth vs. latency;
%(3) adapation --- global view vs. local view; asynchrnous data migration; when to trigger data migration 
%(4) DRAM management  (minimize data movement) --> dynamic programming
%(5) data placement refinement
\vspace{-5pt}

%\textbf{Step 1: phase profiling.} 
\subsubsection{Phase Profiling}
This step collects memory access information for each phase. This information is leveraged by the second and third steps to decide data placement for each phase.
%The workload characteristics should capture the bandwidth consumption of target data objects. 

We rely on hardware performance counters widely deployed in modern processors. 
In particular, we collect the number of last level cache miss event,
%and \textbf{xxx}, 
and then map the event information to data objects. Leveraging the common sampling mode in performance counters (e.g., Precise Event-Based Sampling from Intel or Instruction-based Sampling from AMD),
we collect memory addresses whose associated memory references cause last level cache misses. 
Those memory addresses help us identify target data objects that have frequent memory accesses in main memory.
%are recorded in a pre-allocated buffer accessible by the runtime.

Note that the number of last level cache misses can reflect how intensive main memory accesses happen within a fixed sampling interval.
It works as an indication for which target data objects potentially suffer from the performance limitation of NVM. 
However, there are other events that cause main memory accesses, such as cache line eviction and prefetching operations.
The current performance counters either do not support counting such event (cache line eviction) or do not have the sampling mode
for such event (prefetching operation). Hence, we cannot include those events when counting main memory accesses. %at runtime.
However, the last level cache miss accounts for a large part of main memory accesses.
It can work as a reliable indicator to direct data placement, as shown in the evaluation section. The last level cache miss is also one of the most common events in modern processors, which makes our runtime highly portable across HPC platforms. To compensate for the potential inaccuracy caused by the limitation of performance counters, we introduce constant factors in the performance models in Step 2. 
\vspace{-5pt}
\begin{comment}
    "The stall latency of each access in a modern superscalar processor
    is not the same as the time to access memory. The
    processor can hide the latency to access memory by locating
    multiple memory requests in the instruction stream and
    then using out-of-order execution to issue them in parallel
    to memory via non-blocking caches [39]. In addition, microprocessors
    incorporate prefetchers that locate striding accesses
    in the stream of addresses originating from execution
    and prefetch ahead. As a result the effective latency to access
    memory can be much smaller than the actual physical
    latency for certain access patterns. "
    my comments: "the effects of prefetching is not measured in our HW events,
    because the prefetching does not have instructions.
    streaming access patterns can benefit from prefetching."
\end{comment}

%\textbf{Step 2: performance modeling.}
\subsubsection{Performance Modeling}
Given the memory access information collected for each phase, 
we select those target data objects that have memory accesses 
recorded by performance counters. Those data objects are potential candidates to move from NVM to DRAM.
%However, moving data objects have performance cost. 
To decide which target data objects should be moved,
we introduce lightweight performance models.

\textbf{General description.}
The performance models estimate performance benefit (Equations~\ref{eq:bft_bw} and~\ref{eq:bft_lat})
%(after moving data from NVM to DRAM) 
and data movement cost (Equation~\ref{eq:cost}) between NVM and DRAM. We trigger data movement only when the benefit outweighs the cost. 
To calculate the performance benefit, we must decide if the data object is bandwidth sensitive or latency sensitive (Equation~\ref{eq:app_bw}).
This is necessary to model the performance difference between 
bandwidth sensitive and latency sensitive workloads.

\begin{comment}
In particular, if the data object is bandwidth sensitive, the performance benefit is formulated in Equation~\ref{eq:bft_bw};
if the data object is latency sensitive, the performance benefit
is formulated in Equation~\ref{eq:bft_lat}.
\end{comment}

\textbf{Bandwidth sensitivity vs. latency sensitivity.}
To decide if a target data object in a phase is bandwidth sensitive or latency sensitive, we use Equation~\ref{eq:app_bw}. This equation estimates main memory bandwidth consumption due to memory accesses to the data object ($BW_{data\_obj}$). 

\begin{equation}
\label{eq:app_bw}
\scriptsize
BW_{data\_obj} = \frac{\#data\_access \times cacheline\_size}{\frac{\#samples\_with\_data\_accesses}{\#samples} \times phase\_execution\_time}
\end{equation}

The numerator of Equation~\ref{eq:app_bw} is the accessed data size.
$\#data\_access$ in the numerator is the number of memory accesses to the data object in main memory.
$\#data\_access$ is collected in Step 1 (phase profiling) with performance counters.
\begin{comment}
%We further weight $\#data\_access$ by $\#samples$. 
%$\#samples$ is the number of samples collected for a s
In Step 1, we use a sampling-based approach to collect performance events.
This means that we periodically examine the last cache miss. Hence, in the two equations, $\#data\_access$ is weighted by 
$\#samples$ 
the total number of samples in a 

($\#data\_access \times \#samples$).
$\#samples$ is calculated by $phase\_execution\_time$ divided by $sampling\_interval$.
$phase\_\\execution\_time$ is measured in the profiling iteration.
\end{comment}
For a target data object in a phase, the accessed total data size is
calculated as ($\#data\_access \times cacheline\_size$).
%$\#data\_access \times \#samples \times cacheline\_size$.

The denominator of the equation is the fraction of the execution time that has memory accesses to the target data object in main memory.
This fraction of the execution time is calculated based on $\frac{\#samples\_with\_data\_accesses}{\#samples}$, which is the ratio between the number of samples that collect non-zero accesses to the target data object and the total number of samples.

For example, suppose that the phase execution time is 10 seconds, the hardware counter sampling rate is 1000 cycles, and the CPU frequency is 1 GHz. Then we will have $10^7$ samples in total during the phase execution. Assuming that $10^5$ samples of all samples have memory accesses to the data object, then the fraction of the execution time that accesses the data object is $\frac{10^5}{10^7} \times 10 = 0.1s$. 

Given a data object in a phase, if its  $BW_{data\_obj}$ reaches $t_1$\% of  the peak NVM bandwidth $BW_{peak}$ ($t_1=80$ in our evaluation), then this data object is most likely to be bandwidth sensitive. The performance benefit after moving the data object from NVM to DRAM (i.e., $BFT_{data\_obj\_bw}$) is dominated by the memory bandwidth effect, and can be
calculated based on Equation~\ref{eq:bft_bw}, which will be discussed next.
If $BW_{data\_obj}$ of the data object is less than $t_2$\% of $BW_{peak}$ ($t_2=10$ in our evaluation), then this data object is most likely to be highly latency sensitive.
The performance benefit after moving the data object from NVM to DRAM (i.e., $BFT_{data\_obj\_lat}$) is dominated by the memory latency effect, and can be
calculated based on Equation~\ref{eq:bft_lat}, which will be discussed next.
If $BW_{data\_obj}$ of the data object 
is between $t_1$\% and $t_1$\%, then the data object is likely
to be sensitive to either bandwidth or latency. %or both.
The performance benefit after data movement from NVM to DRAM
is estimated by $max(BFT_{data\_obj\_bw}, BFT_{data\_obj\_lat})$.
To measure $BW_{peak}$, we run 
a highly memory bandwidth intensive benchmark, the STREAM benchmark~\cite{stream_benchmark}, with maximum memory concurrency, and use Equation~\ref{eq:app_bw} and performance counters.

%may have many concurrent memory accesses, but those memory accesses do %not reach the peak  NVM bandwidth. Placing the data object in DRAM 
%does not help performance.

\textbf{Calculation of data movement benefit.}
Equations~\ref{eq:bft_bw} and~\ref{eq:bft_lat} calculate performance benefits (after data movement from NVM to DRAM) for bandwidth sensitive and latency sensitive data objects, respectively. 
The two equations are simply based on an estimation on the performance difference between running the application on NVM and on DRAM.
If the data object is bandwidth-sensitive, then
the application performance on a specific memory is
modeled by $\frac{accssed\_data\_size}{mem\_bw}$ ($mem$ is NVM or DRAM).
%which accounts for memory bandwidth impact.
$accessed\_data\_size$ is \\ $\#data\_access \times cacheline\_size$,
the same as the one in Equation~\ref{eq:app_bw}. 
If the data object is latency-sensitive, then the application
performance on a specific memory is modeled by ${\#data\_access \times mem\_lat}$ ($mem$ is NVM or DRAM). %which accounts for memory latency impact.  
%$\#data\_access$ in the two equations is the number of memory accesses to the data object in main memory.
%$\#data\_access$ is collected in Step 1 with performance counters.
\begin{comment}
%We further weight $\#data\_access$ by $\#samples$. 
%$\#samples$ is the number of samples collected for a s
In Step 1, we use a sampling-based approach to collect performance events.
This means that we periodically examine the last cache miss. Hence, in the two equations, $\#data\_access$ is weighted by 
$\#samples$ 
the total number of samples in a 

($\#data\_access \times \#samples$).
$\#samples$ is calculated by $phase\_execution\_time$ divided by $sampling\_interval$.
$phase\_\\execution\_time$ is measured in the profiling iteration.
\end{comment}
%For a specific data object in a phase, the accessed total data size is calculated as $\#data\_access \times cacheline\_size$.
%$\#data\_access \times \#samples \times cacheline\_size$.

\scriptsize
\begin{multline}
\label{eq:bft_bw}
    BFT_{data\_obj\_bw} = \\ (\frac{\#data\_access \times cacheline\_size}{NVM\_bw} - \\ 
     \frac{\#data\_access \times cacheline\_size}{DRAM\_bw}) \times CF\_bw
\end{multline}

\begin{multline}
\label{eq:bft_lat}
    BFT_{data\_obj\_lat} = \\ (\#data\_access \times NVM\_lat - \\ 
     \#data\_access \times DRAM\_lat) \times CF\_lat
\end{multline}
\normalsize

In the above two equations, we have constant factors $CF\_bw$ (see Equation~\ref{eq:bft_bw}) and $CF\_lat$ (see Equation~\ref{eq:bft_lat}).
Such constant factors are used to improve modeling accuracy.
To meet high performance requirement of our runtime,
the performance models are rather lightweight, and only capture the critical impacts of memory bandwidth or memory latency.
However the models ignore some important performance factors (e.g., overlapping between memory accesses, and overlapping between memory accesses and computation).
Also, the limitation of the sampling-based approach to
count performance events can underestimate the number of memory accesses due to the inability of counting cache eviction and prefetching operations and sampling nature of the approach.
The constant factors $CF\_bw$ and $CF\_lat$ work as a simple but powerful approach to improve modeling accuracy without increasing modeling complexity and runtime overhead.

The basic idea of the two factors is to measure performance ratios between measured performance and predicted performance for representative workloads, and then use the ratios to improve online modeling accuracy for other workloads.

\begin{comment}
In particular, the constant factor $CF\_bw$ is obtained offline. Since Equation~\ref{eq:bft_bw} 
targets on the data object that is sensitive to memory bandwidth,
we run the benchmark STREAM to obtain $CF\_bw$.
In particular, we calculate the performance ratio
between the predicted performance and measured performance 
%when the major data arrays ($a, b$ and $c$) in the benchmark are placed in DRAM
%and other data is placed in NVM.
when running the benchmark in DRAM.
%In particular, we calculate the performance ratio
%between the performance with NVM (no DRAM) and the performance when the 
%major data array in the benchmark is placed in DRAM.
Such performance ratio is $CF\_bw$. The predicted performance is calculated based on ($\#data\_access \times cacheline\_size/DRAM\_bw$), where $\#data\_access$ is collected with performance counters using the sampling-based approach.
Hence, $CF\_bw$ accounts for the potential performance difference between the sampling-based approach and real performance.
\end{comment}

In particular, we run the bandwidth-sensitive benchmark STREAM to obtain $CF\_bw$ offline.
We calculate the performance ratio between the predicted performance and measured performance,
and such ratio is $CF\_bw$. 
The predicted performance is calculated based on ($\#data\_access \times cacheline\_size/DRAM\_bw$), where $\#data\_access$ is collected with performance counters using the sampling-based approach. Hence, $CF\_bw$ accounts for the potential performance difference between our sampling-based modeling and real performance.
The constant factor $CF\_lat$ is obtained in the similar way, except that
we use a latency-sensitive benchmark, the pointer-chasing benchmark~\cite{pointer_chasing} (using a single thread and no concurrent memory accesses).
Also, to calculate the predicted performance, we use ($\#data access \times DRAM\_{lat}$).
Given a hardware platform, $CF\_bw$ and $CF\_lat$ need to be calculated only once.

\textbf{Calculation of data movement cost.}
Data placement comes with data movement cost.
%Equation~\ref{eq:cost} estimates data movement cost $COST_{data\_obj}$.  
The data movement cost can be simply calculated based on data size 
and memory copy bandwidth between NVM and DRAM, which is ($\frac{data\_size}{mem\_copy\_bw}$). 
To reduce the data movement cost, we want to overlap
data movement with application execution. 
This is possible with a helper thread
that runs in parallel with the application to implement
an asynchronous data movement. We discuss this in details 
in Section~\ref{sec:impl}. 
In summary, the data movement cost ($COST_{data\_obj}$) 
is modeled in Equation~\ref{eq:cost}
with the overlapped cost ($mem\_comp\_overlap$) included.

\begin{comment}
To overlap data movement with the application execution, 
we introduce a helper thread that performs data movement (see Section~\ref{sec:impl} for details).
The helper thread runs in parallel with the main program,
which reduces the data movement cost by $mem\_comp\_overlap$.
We describe how to calculate $mem\_comp\_overlap$ as follows.
\end{comment}

\begin{equation}
\label{eq:cost}
\scriptsize
COST_{data\_obj} = max(\frac{data\_size}{mem\_copy\_bw} - mem\_comp\_overlap, 0)
\end{equation}

We describe how to calculate $mem\_comp\_overlap$ as follows.
To minimize the data movement cost, we want to overlap data movement
with application execution as much as possible.
Meanwhile, we must respect data dependency and ensure execution correctness. This means during data movement, the migrated data object must not be read or written by the application.
Given the above requirement on respecting data dependency and minimizing the data movement cost, we can estimate $mem\_comp\_overlap$. 

\begin{comment}
%how to draw a colored circle with text in it.
%see this http://latex-community.org/forum/viewtopic.php?t=25483
%\tikz \node[circle,scale=0.75,color=white, fill=blue]{1}; 
\begin{figure*}
%\centering
\begin{minipage}[t]{.33\textwidth}
  \centering
  \includegraphics[width=\linewidth]{figures/trigger_migration.pdf}
  \caption{An example for determining where to trigger data migration in Step 2. The blue circle is the point to trigger the migration of the data object $a$ for the phase $i$.}
\label{fig:trigger_migration}
\end{minipage}%

\begin{minipage}[t]{.33\textwidth}
  \centering
  \includegraphics[width=\linewidth]{figures/trigger_migration.pdf}
  \caption{Another figure}
  \label{fig:local_vs_global}
\end{minipage}
\end{figure*}
\end{comment}

\begin{comment}
Given a phase $i$ and a data object to be migrated for $i$, we trigger the movement of this data object in the beginning of a preceding phase $j$. Between the phases $j$ (including the phase $j$) and $i$, the data object is not referenced, while the immediately preceding phase of the phase $j$ references the data object. $mem\_comp\_overlap$ is the application execution time between the phases $j$ and $i$. 
\end{comment}

\begin{figure}
%\centering
  \centering
  \includegraphics[height=0.2\textheight, width=0.3\textwidth]{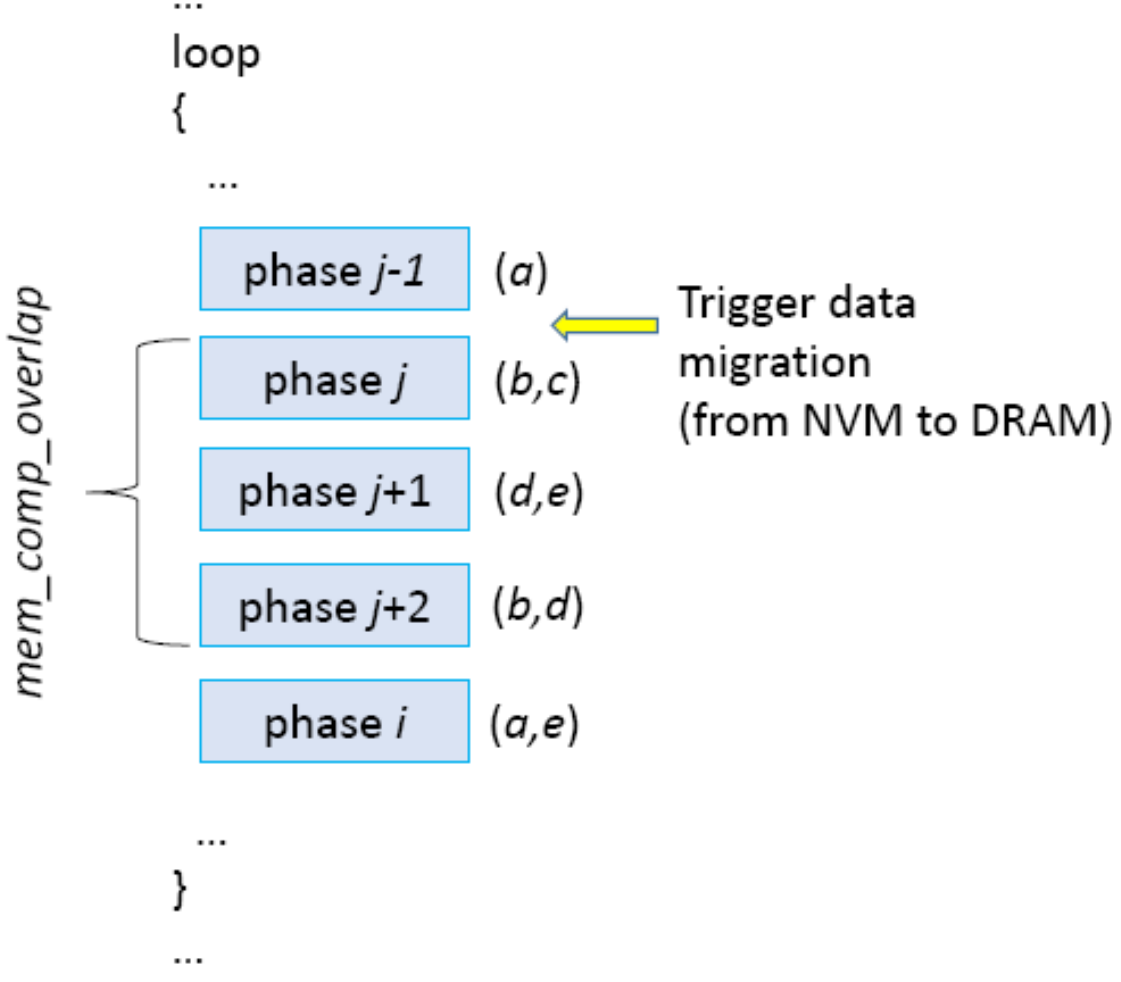}
  \vspace{-10pt}
  \caption{An example to show how to calculate $mem\_comp\_overlap$  for the data object $a$ in the phase $i$. The yellow arrow is the point to trigger the migration of $a$ from NVM to DRAM for the phase $i$, if $a$ is not in DRAM. The letters in brackets represent target data objects referenced in the corresponding phases.}
\label{fig:trigger_migration}
\vspace{-15pt}
\end{figure}

Figure~\ref{fig:trigger_migration} explains how to calculate 
$mem\_comp\_overlap$ with an example. This example shows how to calculate
$mem\_comp\_overlap$ for a data object ($a$)  in a specific phase (the phase $i$). 
If $a$ is not in DRAM, we can trigger data migration of $a$ as early as the beginning of the phase $j$, because $a$ is not referenced between $j$ and $i$. 
We cannot trigger data migration of $a$
at the beginning of the phase $j-1$, because $a$ is referenced there. $mem\_comp\_overlap$ is the application execution time between the phases $j$ and $i$. The data movement time, $\frac{data\_size}{mem\_copy\_bw}$, can
be smaller than $mem\_comp\_overlap$. In this case, the data movement
is completely overlapped with application execution, and the data movement cost $COST_{data\_ obj}$ is 0.

Our estimation on $COST_{data\_obj}$ could be an overestimation (a conservative estimation). In particular,
when a data object is to be migrated from NVM to DRAM for a phase,
it is possible that the data object is already in DRAM.
Use Figure~\ref{fig:trigger_migration} as an example again.
Since the phase $j-1$ references $a$, it is possible that
$a$ is already in DRAM before the point to trigger the data migration.
Also, $COST_{data\_obj}$ does not include the cost of moving data from 
DRAM to NVM when there is no enough space in DRAM and we need to switch data.
Such overestimation and ignorance of data movement from DRAM to NVM are due to the fact that the data movement cost
for each phase is isolatedly calculated during the modeling time.
Hence, what data objects are in DRAM and whether there is enough space in DRAM is uncertain during the modeling time. We will solve the above problems in the next step (Step 3). 

\begin{comment}
The above discussion on $COST_{data\_obj}$ assumes that there is enough empty memory space in DRAM to move the data object from NVM to DRAM. If there is no enough space, then some existing data objects in DRAM must be moved from DRAM to NVM. Such data movement cost should be included in $COST_{data\_obj}$. However, during the modeling time, whether 
there is enough space in DRAM is uncertain, because the data placement plan is not determined yet.
\end{comment}

\begin{comment}
For such case, $COST_{data\_obj}$ should include the data movement cost from DRAM to NVM. To calculate such cost, we must decide which data object in DRAM must be moved from DRAM to NVM. We make such decision based on the sizes of data objects in DRAM. In particular, we move data objects from DRAM to NVM whose total size is just big enough to allow the target data object to move from NVM to DRAM. Furthermore, the data movement from DRAM to NVM should be overlapped with application execution as much as possible. Using the same approach as the above discussion on the data movement from NVM to DRAM, we proactively trigger the data movement from DRAM to NVM and include
this cost in $COST_{data\_obj}$. 
\end{comment}

\begin{comment}
Based on the above formulation and discussion, we can calculate the benefit and cost of data movement from NVM to DRAM for any data object referenced in a specific phase.
\end{comment}

%Hence, the data movement cost can be zero (i.e., completely hidden).
%We will discuss $mem\_comp\_overlap$ in details in \textbf{Step x}.

%\textbf{Step 3: data placement decision.}
\subsubsection{Data Placement Decision and Enforcement}
\label{sec:dp_decision_enforcement}
Based on the above formulation for the benefit and cost of data movement, we determine data placement for all phases \textit{one by one}. In particular, to determine data placement for a specific phase, 
we define a weight $w$ for each target data object referenced in this phase:

\begin{equation}
w = BFT_{data\_obj} - COST_{data\_obj} - extra\_COST_{data\_obj}
\end{equation}

$extra\_COST_{data\_obj}$ accounts for the data movement cost, 
when there is no enough space in DRAM to move the target data object
from NVM to DRAM and we have to move data from DRAM to NVM to save space.
To calculate $extra\_COST_{data\_obj}$, we must decide which data object in DRAM must be moved. We make such decision based on the sizes of data objects in DRAM. In particular, we move data objects from DRAM to NVM whose total size is just big enough to allow the target data object to move from NVM to DRAM. 
Note that since we determine data placements for all phases one by one,  when we decide the data placement for a specific phase, we have made the data placement decisions for previous phases. Hence, we have a clear knowledge on which data objects are in DRAM and whether the target data object is already in DRAM.
%Based on the knowledge, it is also possible to know whether the target data object is already in DRAM. If it is, then there will be no data movement cost.

%Furthermore, the data movement from DRAM to NVM should be overlapped with application execution as much as possible. 

Besides the weight $w$, each data object has a data size.
Given the DRAM size limitation, our data placement problem
is to maximize total weights of data objects in DRAM while
satisfying the DRAM size constraint.
This is a 0-1 knapsack problem~\cite{knapsackbook}.

The knapsack problem can typically be solved by dynamic programming
in pseudo-polynomial time. If each data object has a distinct value per unit of weight ($data\_size/w$),
the empirical complexity is $O((log(n))^2)$~\cite{knapsackbook}, where $n$ is the number of target data objects referenced in a phase.

The above approach can determine data placement
for individual phases. We name this approach as ``phase local search''. Determining data placement at the granularity of individual phases can lead to the optimal data placement for each phase, but result in frequent data movements, some of which may not be able to be completely overlapped by application execution. Alternatively, determining data placement at the granularity of all phases (named ``cross-phase global search'') has less data movement than phase local search, because all phases are in fact treated as a combined single phase: Once the optimal data placement is determined within the combination of all phases, there is no data movement within the combination. However, the optimal data placement for the combination of all phases does not necessarily result in the best performance for each individual phase.

%\textbf{Dong: add more details about the dynamic programming. formulate global search and local search}

\begin{comment}
as well as a combination of all phases.
For a combination of all phases, once the optimal data placement is determined,
there is no data movement within the combination. Hence,
treating all phases as a combination, we minimize the data movement.
However, the optimal data placement for the combination of all phases does not necessarily result in the best performance for each individual phase.
Determining the data placement at the granularity of single phase
can lead to the optimal data placement for each phase, but result in more frequent data movement. If those extra data movement cannot be hidden,
then determining the data placement at the granularity of single phase (called ``local view'') may have worse performance than determining the data placement at the granularity of all phases (called ``global view'').
\end{comment}

Based on the above discussion, we use dynamic programming to
\textit{determine the data placement using both phase local search and cross-phase global search}, and then choose the best data placement of the two searches.

%\textbf{Step 4: data placement enforcement.}
%proactive data movement --> the data movement may disturb the application execution.
After we make the data placement decision at the end of the first iteration, we enforce data placement since the second iteration.
At the beginning of each phase, the runtime asks a helper thread (see Section~\ref{sec:impl} for implementation details)
to proactively move data objects between NVM and DRAM based on the data placement decision for future phases.

\begin{figure*}
%\centering
  \centering
  \includegraphics[height=0.18\textheight, width=0.7\textwidth]{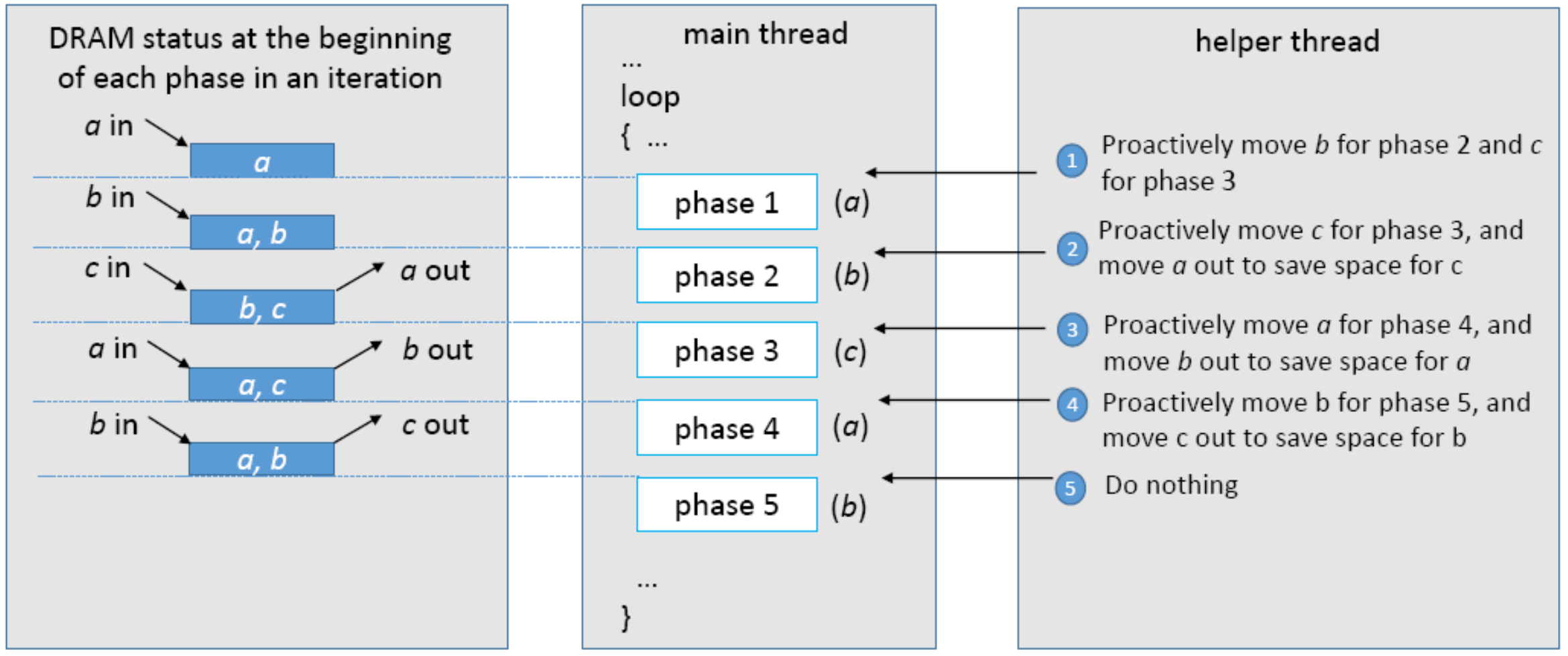}
  \caption{An example to show proactive data migration with a helper thread. The letters in the figure represent data objects. The letters in brackets (e.g., (a) and (b)) represent target data objects that are determined to be placed in DRAM for the corresponding phases. DRAM can hold two data objects at most.}
\label{fig:proactive_migration}
\end{figure*}

Figure~\ref{fig:proactive_migration} gives an example for how to enforce data placement with a helper thread after determining data placement. In this example, there are three target data objects ($a, b$, and $c$) and five phases. The data placement decision for each phase
is represented with letters in brackets (e.g., ($a$) for the phase 1). 
We assume DRAM can hold two data objects at most. 
The data movement enforced by the helper thread respects data dependence across phases and the availability of DRAM space. Such example is a case of phase local search, where each phase makes 
its own decision for data placement. There are eight data movements in total. With a cross-phase global search, only two data objects will be moved to DRAM for all phases. %because they are the most common ones in this example. 
The cross-phase global search results in only two data movements.
Based on the performance modeling and dynamic programming, we can decide whether the cross-phase global search or phase local search is better. 
\vspace{-10pt}

\subsection{Optimization}
\label{sec:opt}
To improve runtime performance, we introduce a couple of optimization techniques as follows.

\textbf{Handling workload variation across iterations.}
In many scientific applications, the computation and memory access patterns remain stable across iterations.
This means once the data placement decision is made at the end of the first iteration, 
we can reuse the same decision in the rest of iterations.
However, some scientific applications have workload variation across iterations.
We must adjust data placement decision correspondingly.

To accommodate workload variation across iterations, Unimem monitors the performance of each phase after data movement. If there is obvious performance variation (larger than 10\%),
then the runtime will activate phase profiling again and adjust the data placement decision. 
%Such performance monitor is based on, hence it is very lightweight. 

\textbf{Initial data placement.} 
By default, all data objects are initially placed in NVM and moved between DRAM and NVM by Unimem at runtime.
However, data movement can be expensive, especially for large data objects, even though we use the proactive data movement to overlap data movement
with application execution.
%In fact, comparing with our lightweight performance models and execution profiling, the data movement accounts for most of the runtime overhead. 
To reduce the data movement cost, we selectively
place some data objects in DRAM at the beginning of the application, instead of placing all data objects in NVM. 
The existing work has demonstrated performance benefit of the initial data placement on GPU with HMS~\cite{asplos15:agarwal, pcm_gpu_pact13}.
Our initial data placement technique on NVM-based HMS is consistent with those existing efforts.
%This method removes some of the data movement off the critical path.

%To determine which data objects should be placed in DRAM initially, 
For initial data placement, we place in DRAM those target data objects with the largest amount of memory references (subject to the DRAM space limitation).
To calculate the number of memory reference for each target data object, we employ compiler analysis and represent the number of memory reference
as a symbolic formula with unknown application information, similar to~\cite{icpcds99:ding}.
Such information includes the number of iterations and coefficients of array access. This information is typically available before the main computation loop and before memory allocation for target data objects.
Hence it is possible to decide and implement initial data placement before main computation loop for many HPC applications. 
%In particular, we employ static analysis to count the total amount of memory accesses for each 
%target data object.
However, we cannot determine initial data placement for those data objects that do not have the information available before the main computation loop (e.g., the number of iteration is determined by a convergence test at run time).

Our method determines initial data placement simply based on the number of memory reference and ignores caching effects.
The ignorance of caching effects can impact the effectiveness of initial data placement.
In particular, some data objects with intensive memory references may have good reference locality and do not cause a lot of main memory accesses.
However, our practice shows that in all cases of our evaluation, initial data placement based on compiler analysis makes the data placement decision consistent with the runtime data placement decision using the cross-phase global search.
Using compiler analysis can work as a practical and effective solution to direct initial data placement, because the target data objects with a large amount of memory references tend to frequently access main memory.

\textbf{Handling large data objects.} 
We move data between DRAM and NVM at the granularity of data object. This means a data object larger than the DRAM space cannot be migrated.
This problem is common to any software-based data management on HMS.

A method to address the above problem is to partition the large data object into multiple chunks with each chunk smaller than the DRAM size.
At runtime, we can profile memory access for each chunk instead of the whole data object, and move data chunk if the benefit overweight the cost of data chunk movement.
This method exposes new opportunities to manage data and improve performance.

However, this solution is not always feasible, because it can involve a lot of programming efforts to refactor the application such that memory references to the large data object are based on chunk-based partitioning.
A compiler tool can be helpful to transform some regular memory references into new ones based on chunk-based partitioning (assuming the input problem size and number of loop iterations are known). However, this kind of automatic code transformation can be impotent for high-dimensional arrays with the notorious memory alias problem and irregular memory access patterns. In Unimem, we employ a conservative approach which only partitions those one-dimensional arrays with regular memory references. %and transform the code correspondingly.

In our evaluation with representative numerical kernels, we find that partitioning large data objects is often not helpful, %because the memory referecnes to multiple chunks often happen in one phase 
because making the data placement decision based on chunks leads to much more frequent data movements,
most of which are difficult to be overlapped with application execution and hence exposed to the critical path, but we do have a benchmark (FT) benefit from partitioning large data objects.

%In general, we expect that partitioning large data objects is a conservative optimization technique, and may take effects in limited applications.

\begin{comment}
\textbf{Phase combination.}
Some phase can be small in terms of execution time.
For such phase, the phase profiling and performance modeling alone can take most of the phase execution time.
This kind of phase is commonly spotted in communication intensive applications. To amortize the phase profiling and performance modeling cost, we do not consider such small phases.
Instead, those phases are combined with neighbor phases to build a larger phase. We make phase combination when the phase profiling and performance modeling of a phase take more than 20\% of the phase execution time.
\end{comment}

\subsection{Implementation}
\label{sec:impl}
We have implemented Unimem as a runtime library to perform online adaptation of data placement on HMS. To leverage the library, the programmer needs to insert a couple of APIs into the application.
Such change to the application is very limited, and is used to initialize the library and identify the main computation loop and target data objects.
In all applications we evaluated, the modification to the applications is less than 20 lines of code.
Table~\ref{tab:api} list those APIs and their functionality.

%``We target parallel applications from the HPC domain with
%iterative structure, such that each program phase is executed many times.
%We exploit this property to collect hardware event rates during
%the first few executions of each phase to serve as input for the
%model. We hardcode the model itself into the runtime system by
%programming the coefficients derived during the training process
%for a particular model into the library. The runtime system facilitates
%online predictions of performance based on the collected
%hardware event rates. ''

\begin{figure}
	\vspace*{10pt}
    \centering
    \includegraphics[width=0.48\textwidth, height=0.18\textheight]{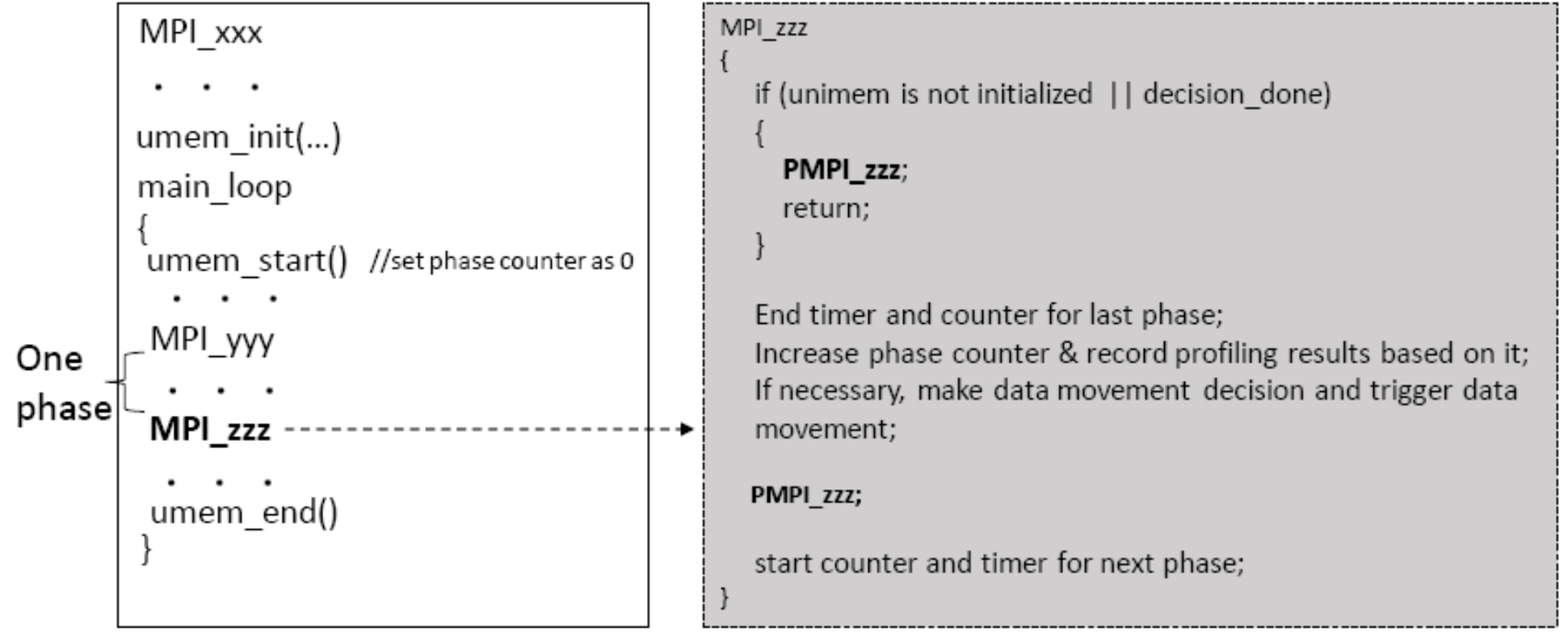}
    \caption{Transparently identifying phases based on PMPI.}
    \label{fig:pmpi}
    \vspace{-20pt}
\end{figure}

\begin{table}
        \begin{center}
        \vspace{-10pt}
        \caption{APIs for using Unimem runtime}
        \vspace{-10pt}
        \label{tab:api}
        \scriptsize
        \begin{tabular}{|p{2.3cm}|p{4.3cm}|}
        \hline
        \textbf{API Name}     & \textbf{Functionality}                                 \\ \hline \hline
        {\fontfamily{qcr}\selectfont unimem\_init}       & initialization for hardware counters, timers and global variables            \\\hline
        {\fontfamily{qcr}\selectfont unimem\_start}      & identify the beginning of the main computation loop                 \\ \hline
        {\fontfamily{qcr}\selectfont unimem\_end}       & identify the end of the main computation loop                    \\ \hline
        {\fontfamily{qcr}\selectfont unimem\_malloc}    & identify and allocate target data objects                    \\ \hline
        {\fontfamily{qcr}\selectfont unimem\_free}   & free memory allocation for target data objects                  \\ \hline
        \end{tabular}
        \end{center}
        \vspace{-20pt}
\end{table}
The runtime library decides data placement at the granularity of execution phase. As discussed before, a phase is delineated by MPI operations.
To automatically form phases, we employ the MPI standard profiling interface (PMPI). 
{\fontfamily{qcr}\selectfont PMPI\_} function behaves in the same way as 
{\fontfamily{qcr}\selectfont MPI\_} function, but PMPI allows one 
to write functions that have the behavior of the standard function plus any other behavior one would like to add.
Based on PMPI, we can transparently identify execution phases and control profiling without programmer intervention.
Figure~\ref{fig:pmpi} depicts the general idea. In particular,
we implement an MPI wrapper based on PMPI. The wrapper encapsulates 
the functionality of enabling and disabling profiling and uses a global
counter to identify phases. 
%the into MPI collective operation, blocking communication and synchronization 

To identify target data objects, the programmer must use \\{\fontfamily{qcr}\selectfont unimem\_malloc} to allocate them before the main computation loop.
This API allows Unimem to collect pointers pointing to target data objects.
Collecting those pointers are necessary to implement data movement without asking the programmer to change the application after data movement. 
%the runtime collects pointers pointing to the target data objects in {\fontfamily{qcr}\selectfont umem\_init} 
%(the programmer provides those pointers).
In particular, after data movement for a target data object, the runtime changes the data object pointer and makes it point to the new memory space of the data object without disturbing execution correctness. If there is a memory alias to the data object but such alias is created within the main computation loop, then the memory alias can still work correctly, because it is updated in each iteration and will point to the new memory space of the data object after data movement.
If the memory alias to the data object is created before the main computation loop, then such memory alias information must be explicitly sent to the runtime by the programmer using {\fontfamily{qcr}\selectfont unimem\_malloc}, such that the memory alias can be updated and points to the correct memory space after data movement.
%due to the iterative structure of the main computation loop. 

The DRAM space is limited in HMS. To manage the DRAM space, we avoid making any change to the operating system (OS), and introduce
a user-level service. Each node runs an instance of such service.
The service coordinates the DRAM allocation from multiple MPI processes on the same node.
In particular, the service responds to any DRAM allocation request from the runtime, and bounds
the memory allocation within the DRAM space allowance.
Our current implementation for such service 
is based on a simple memory allocator without consideration of memory allocation efficiency and fragmentation,
because we expect that data movement should not be frequent, and data allocation for data movement should not be frequent for performance reason.
However, an advanced implementation could be based on an existing memory allocator, such as HOARD~\cite{asplos00:berger} and the lock-free allocator~\cite{pldi04:maged}.
%Furthermore, to transparently implement the interaction between the service and MPI processes,
%we track an application's memory allocation by using features of the GNU linker to interpose the interaction code between the application and memory allocation functions (e.g., malloc, free, new, and delete).

As discussed in Section~\ref{sec:design} (see Step 2), we use a helper thread to proactively trigger data movement, such that data movement 
%is not in the critical path and 
is overlapped with  application execution. The helper thread is invoked in {\fontfamily{qcr}\selectfont unimem\_init}. In the main computation loop, the helper thread and the main thread interact through a shared FIFO queue. The main thread puts data movement requests into the queue; the helper thread checks the queue, performs data movement, and removes the data movement request off the queue once the data movement is done. 
At the beginning of each phase, the runtime of the main thread will check the queue status to determine if all proactive data movement for the current phase is done. Hence, the queue works as a synchronization mechanism between the helper thread and the main thread. 
Note that checking the queue status and putting data movement requests into the queue is lightweight, because we avoid frequent data movement in our design. 

\begin{figure}
   \centering
   \includegraphics[height=0.15\textheight]{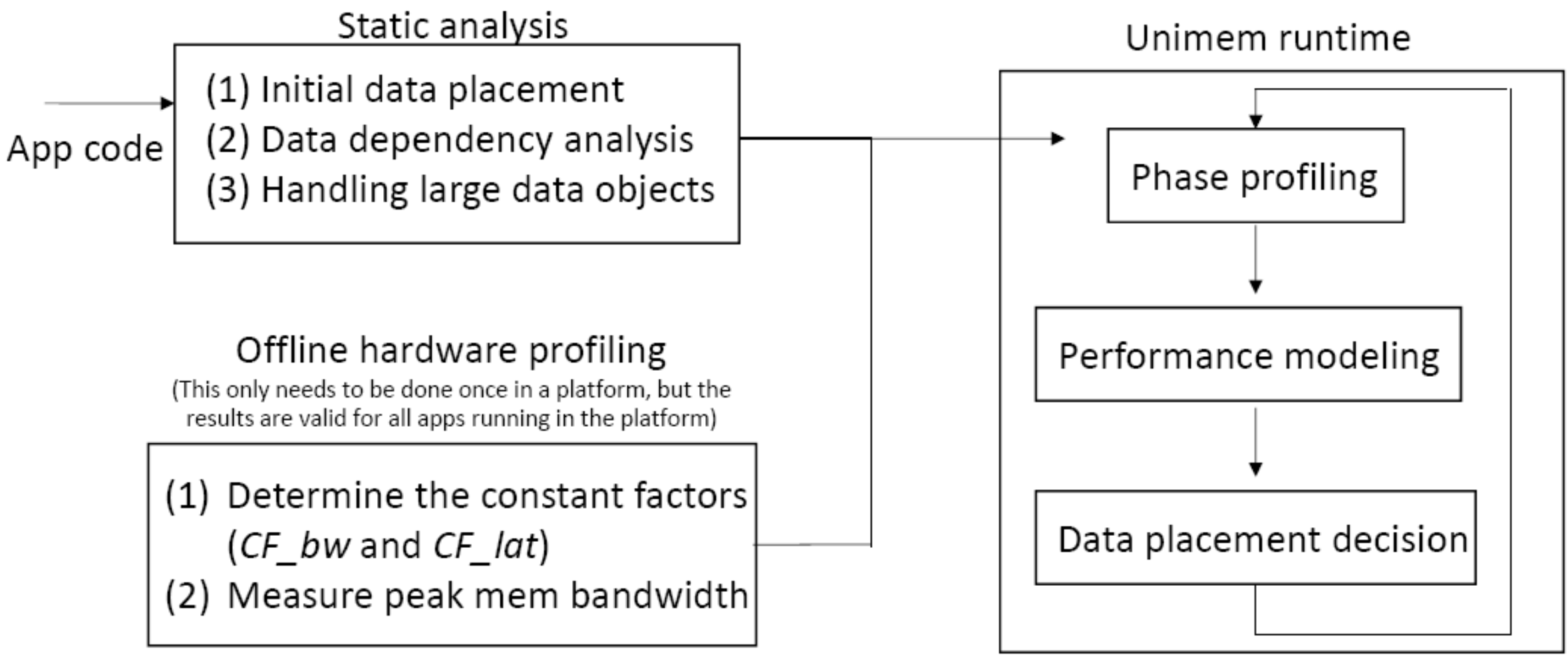}
   \vspace{-10pt}
   \caption{The general workflow for Unimem.}
   \label{fig:general_workflow}
   \vspace{-20pt}
   	\vspace*{5pt}
\end{figure}

As discussed in Section~\ref{sec:design} (see Step 2), to ensure execution correctness, the runtime must respect data dependency across phases when moving data objects with the helper thread.
The data dependency check is implemented by static analysis.
We introduce an LLVM~\cite{Lattner:Mthesis} pass to analyze data references to target data objects 
between MPI calls. 
To handle those unresolved control flows during the static analysis, we
embed data dependency analysis result for each branch,
and delay data dependency analysis until runtime.
The compiler-based data dependency analysis can be conservative
due to the challenge of pointer analysis~\cite{popl03:chakaravarthy}. 
There is also a large body of research related to the
approximation of pointer analysis to improve compiler-based data dependency analysis. However, to simplify our implementation, we currently use a directive-based
approach that allows the programmer to use directives to explicitly inform
the runtime of data dependency for target data objects across phases.
This approach is inspired by task dependency clauses in OpenMP, and works
as a practical solution to address complicated data dependency analysis.
Figure~\ref{fig:general_workflow} depicts the general workflow. %of Unimem.
\vspace{-5pt}

%How to choose the target data objects;
%"we track an application's memory allocation by using features of the GNU linker to interpose 
%code between the application and standard C/C++ memory allocation functions (e.g., malloc, free, new, and delete)"

%MEM\_TRANS\_RETIRED:LOAD\_LATENCY:ldlat=42.  \\

%%%%%%%%%%%%%%%%%%%%%%%%%%%%%%%%%%%%%%%%%%%%%%%%%%%%%%%%%%%%%%%%%%%%%%%%%%%%%%%%%%%%
\begin{comment}
"The stall latency of each access in a modern superscalar processor
is not the same as the time to access memory. The
processor can hide the latency to access memory by locating
multiple memory requests in the instruction stream and
then using out-of-order execution to issue them in parallel
to memory via non-blocking caches [39]. In addition, microprocessors
incorporate prefetchers that locate striding accesses
in the stream of addresses originating from execution
and prefetch ahead. As a result the effective latency to access
memory can be much smaller than the actual physical
latency for certain access patterns.
"

my comments: "the effects of prefetching is not measured in our HW events,
because the prefetching does not have instructions.
streaming access patterns can benefit from prefetching."
\end{comment}
%%%%%%%%%%%%%%%%%%%%%%%%%%%%%%%%%%%%%%%%%%%%%%%%%%%%%%%%%%%%%%%%%%%%%%%%%%%%%%%%%%%%

\section{Evaluation Methodology}
\label{sec:eval_method}
In our evaluation, we use Quartz emulator~\cite{middleware15:volos}.
Quartz enables an efficient emulation of a range
of NVM latency and bandwidth characteristics. 
%for performance evaluation of NVM and their impact
%on applications performance.
%(without modifying or instrumenting
%their source code) by leveraging features available in commodity hardware.
Quartz has low overhead and good accuracy (with emulation
errors 0.2\% - 9\%)~\cite{middleware15:volos}.
We do not use cycle-accurate architecture simulators because of
their slow simulation which cannot scale to large workloads.
Furthermore, Quartz allows us to consider cache eviction effects,
memory-level parallelism, and system-wise memory traffic, which is not available in other state-of-the-art, software-based emulation approaches~\cite{pcmsim, mnemosyne_asplos11}.
However, due to the limitation of Quartz, 
%we cannot distinguish  the latency difference between read and write operations;
we can only emulate either bandwidth limitation or latency limitation,
but cannot emulate both of them.

Using Quartz requires the user to have privilege access to the test system.
We do not have such privilege access on the test platform for our strong scaling tests.
Hence, instead of using Quartz, we leverage NUMA architecture to emulate NVM.
In particular, we carefully manage data placement at the user level, 
such that, given an MPI task, a remote NUMA memory node works as NVM while the NUMA node local to the MPI task works as DRAM.
The latency and bandwidth difference between the remote and local NUMA memory nodes emulates that between NVM and DRAM.
On our test platform for strong scaling tests, the emulated NVM has 60\%
of DRAM bandwidth and 1.89x of DRAM latency.

We have two test platforms for performance evaluation.
One test platform (named ``Platform A'') is a small cluster. Each node of it has two eight-core Intel Xeon E5-2630 processors (2.4 GHz) and 
32GB DDR4. We use this platform for tests in all figures except Figure~\ref{fig:strong_scaling_cg}. %and~\ref{fig:strong_scaling_bt}. 
We deploy Quartz on such platform.
The other test platform is the Edison supercomputer at Lawrence Berkeley National Lab (LBNL). We use this platform for tests in Figure~\ref{fig:strong_scaling_cg}. %and ~\ref{fig:strong_scaling_bt}.
Each Edison node has two 12-core Intel Ivy Bridge processor (2.4 GHz) with 64GB DDR3. 
As discussed before, we perform strong scaling tests and leverage NUMA architecture to emulate NVM on this system. 

We use six benchmarks from NAS parallel benchmark (NPB) suite 3.3.1,
and one production scientific code Nek5000~\cite{27-nek5000}.
For Nek5000, we use eddy input problem with a $256\times256$ mesh.
%\textbf{TODO: which are our target data objects?}
The target data objects of those benchmarks are listed in Table~\ref{tab:data_objects_list}.
Those data objects are the most critical data objects accounting for more than 95\% of
memory footprint except CG and Nek5000. For CG, there are three large data objects ($aelt$, $acol$, and $arow$) only used for problem initialization. They are not treated as target data objects.
For Nek5000, we use main simulation variables and geometry arrays in Nek5000 core. Those are the most important data objects for Nek5000 simulation.
We use GNU compiler (4.4.7 on Platform A and 6.1.0 on Edison) and use default compiler options for building benchmarks. We use the sampling-based approach to collect performance events on the two
platforms. The sampling interval is chosen as 1000 CPU cycles, such that the sampling overhead is ignorable while the sampling is not sparse to lose modeling accuracy. 
%\textbf{TODO}: performance counter sampling interval  is 1000 cycles.
\vspace{-10pt}

\begin{table}
        \begin{center}
        \caption{Target data objects in NPB benchmarks and Nek5000}
        \vspace{-10pt}
        \label{tab:data_objects_list}
        \tiny
        \begin{tabular}{|p{1.2cm}|p{4cm}|p{2cm}|}
        \hline
        \textbf{Benchmark}    & \textbf{Target data objects}  &\textbf{\% of total app mem footprint}                                 \\ \hline \hline
         CG &  $colidx$, $a$, $w$, $z$, $p$, $q$, $r$, $rowst$, $x$ &  42\%  \\ \hline
         FT & $u$, $u0$, $u1$, $u2$, $twiddle$ & 99\%    \\ \hline
         BT & $rhs$, $forcing$, $u$, $us$, $vs$, $ws$, $qs$, $rho\_i$, $square$, $out\_buffer$, $in\_buffer$, $fjac$, $njac$, $lhsa$, $lhsb$, $lhsc$ &  99\% \\ \hline 
         LU & $u$, $rsd$, $frct$, $flux$, $a$, $b$, $c$, $d$, $buf$, $buf1$  &  99\%  \\ \hline
         SP & $u$, $us$, $vs$, $ws$, $qs$, $rho\_i$, $square$, $rhs$, $forcing$, $out\_buffer$, $in\_buffer$, $lhs$ & 98\%  \\ \hline
         MG & $buff$, $u$, $v$, $r$ & 99\%					\\ \hline
         Nek5000(eddy) &  Geometry arrays and main simulation variables (48 data objects in total) & 35\% \\ \hline
        \end{tabular}
        \end{center}
        \vspace{-10pt}
\end{table}

\vspace*{5pt}
\section{Evaluation}
\label{sec:eva}
The goal of our evaluation is multiple-folding. 
First, we want to test if our runtime can effectively direct data placement
to narrow the performance gap between NVM and DRAM;
Second, we want to test if our runtime is lightweight enough;
Third, we want to test the performance of our runtime 
in various system configurations, including different DRAM sizes
and different system scales.
Unless indicated otherwise, performance in this section is normalized to 
that of the DRAM-only system.

\begin{figure}[!t]
    \centering
    \includegraphics[width=0.48\textwidth, height=0.13\textheight]{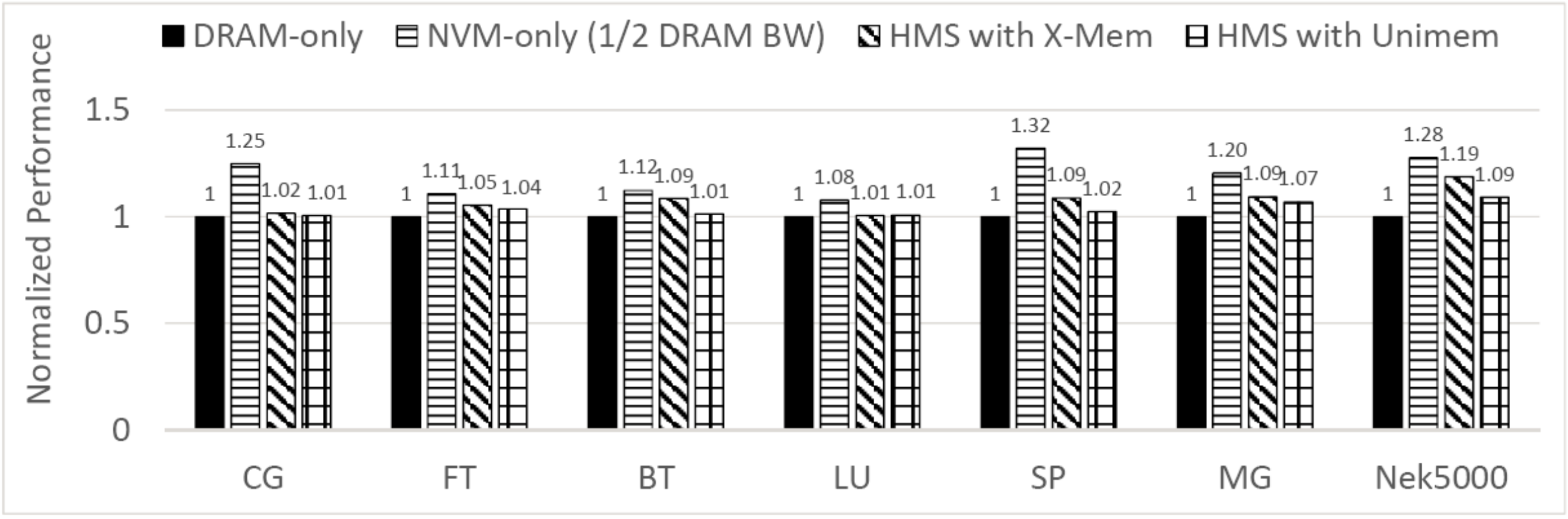}
     \vspace{-20pt}
    \caption{The performance (execution time) comparison between DRAM-only, NVM-only, the existing work (X-Men), and HMS with Unimem. NVM has 1/2 DRAM bandwidth.}
    \label{fig:eval_basic_perf_half_dram_bw}
    \vspace{-10pt}
\end{figure}

\begin{figure}[!t]
    \centering
    \includegraphics[width=0.48\textwidth, height=0.13\textheight]{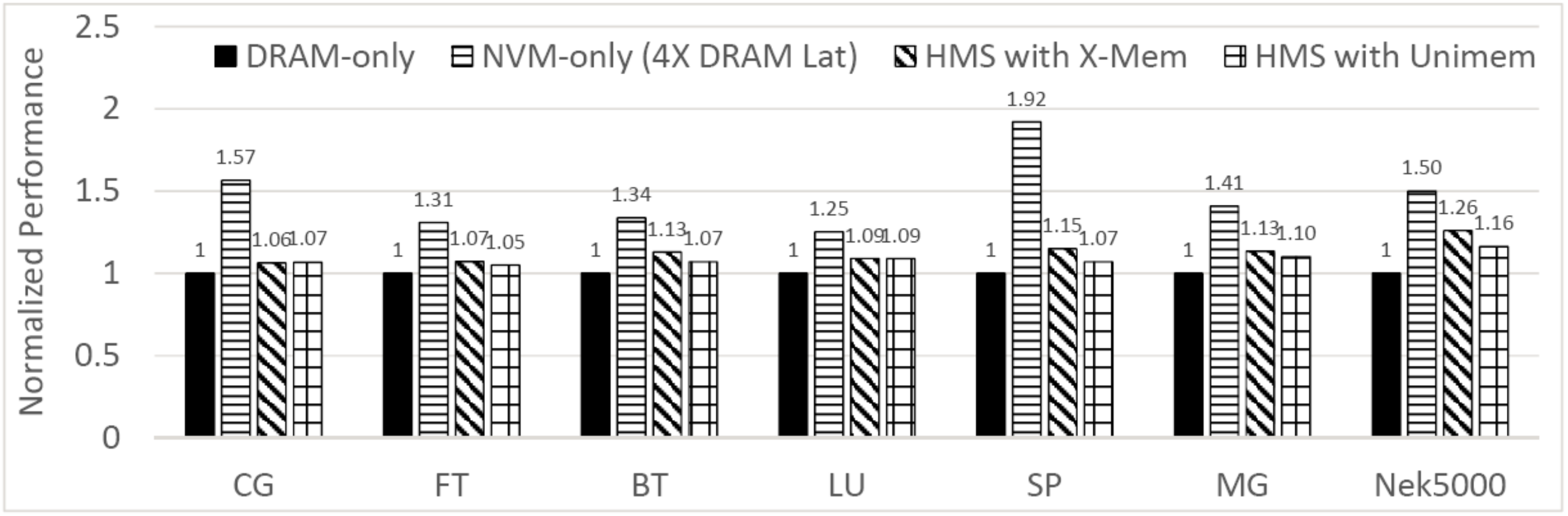}
     \vspace{-20pt}
    \caption{The performance (execution time) comparison between DRAM-only, NVM-only, the existing work (X-Men), and HMS with Unimem. NVM has 4x DRAM latency.}
    \label{fig:eval_basic_perf_four_dram_lat}
    \vspace{-10pt}
\end{figure}

\textbf{Basic performance tests.} 
We compare the performance (execution time) of DRAM-only, NVM-only, and HMS with Unimem. We use four nodes in Platform A with one MPI task per node. 
We use CLASS C as input problem for NPB benchmarks. NVM and DRAM sizes are 16GB and 256MB respectively.
Figures~\ref{fig:eval_basic_perf_half_dram_bw} and~\ref{fig:eval_basic_perf_four_dram_lat} show the results. 
NVM is configured with 1/2 DRAM bandwidth (Figure~\ref{fig:eval_basic_perf_half_dram_bw}) or 4x DRAM latency (Figure~\ref{fig:eval_basic_perf_four_dram_lat}). 

\begin{figure}[!t]
    \centering
    \includegraphics[width=0.48\textwidth, height=0.13\textheight]{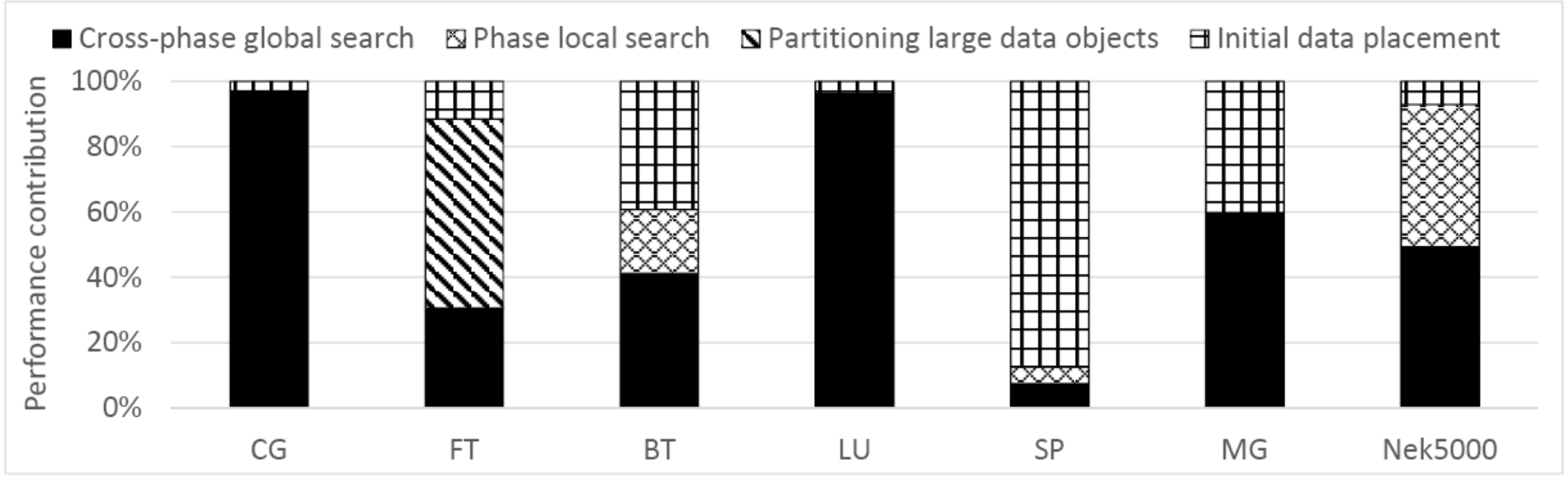}
     \vspace{-20pt}
    \caption{Quantifying the contributions of our four major techniques to performance improvement.}
    \label{fig:eval_basic_op}
    \vspace{-10pt}
\end{figure}

We first notice that there is a big performance gap between NVM-only
and DRAM-only cases. On average, the gap is 18\% for NVM with 1/2 DRAM bandwidth and 47\% for NVM with 4x DRAM latency. %(see ``Overall'' in the figures). 
However, Unimem greatly narrows the gap and makes performance very
close to DRAM-only cases: the average performance difference between DRAM-only and HMS is only 3\% for NVM with 1/2 DRAM bandwidth and 7\% for NVM with 4x DRAM latency, and the performance difference is no bigger than 10\%
in all cases. This demonstrates that Unimem successfully directs
data placement for those performance-critical data objects.
This also demonstrates that Unimem is very lightweight
after we optimize runtime performance and hide data movement cost.

We compare Unimem and X-Men~\cite{eurosys16:dulloor} (a recent software-based solution for data placement in HMS). The results are shown in Figures~\ref{fig:eval_basic_perf_half_dram_bw} and~\ref{fig:eval_basic_perf_four_dram_lat}. 
X-Men uses PIN-based \textit{offline profiling} to characterize memory access patterns and make the decision on data placement. They do not consider data movement cost and assume a homogeneous memory access pattern within a data object. %static nature,
The results show that Unimem performs similarly to X-Men, but performs 10\%
better than X-Men for Nek5000. Nek5000 is a production code with various memory
access patterns across phases. Unimem adapts to those variations, hence performing better. Also Unimem does not need any offline profiling for applications.

\textbf{Detailed performance analysis}. Based on the results of basic performance tests,  we further quantify the contributions of our runtime techniques to performance improvement on HMS. This quantification study
is important to investigate how effective our techniques are and when
they can be effective.
We study four major techniques: (1) cross-phase global search,
%This technique includes using a global view to examine memory access patterns across phases and using a helper thread to overlap data movement and computation. 
(2) phase local search, 
%Using this technique, we determine data placement at the granularity of individual phase to optimize data placement.
(3) partitioning large data objects, 
%We use this technique to expose more opportunities for data placement and performance improvement.
and (4) initial data placement. 
%We use this technique to reduce data movement cost at runtime.

We apply the four techniques one by one. In particular, we apply (1),
and then apply (2) to (1), and then apply (3) to (1)+(2), and then apply (4) to (1)+(2)+(3). We measure the performance variation after
applying each technique to quantify the contribution of each technique
to performance. 
We use the same system configurations as basic performance tests with NVM 
configured with 1/2 DRAM bandwidth.
Figure~\ref{fig:eval_basic_op} shows the results.

We notice that cross-phase global search can be very effective.
In fact, in benchmarks CG and LU, 
more than 90\% of the contribution comes from this technique.
However, cross-phase global search could lose some opportunities to improve performance on individual phases, because it uses the same data placement decision on all phases. Using phase local search can complement cross-phase global search. For BT and SP, using phase local search we improve performance 
by 19\% and 5\% respectively. 

Initial data placement is very useful. In fact, it takes effect on all benchmarks. For SP, it is the most effective approach (87\% contribution comes from this technique). 
%Note that cross-phase global search and phase local search cannot work very well for SP, because the performance event collection based on the sampling mode introduces high inaccuracy to catch memory access patterns. 

Partitioning large data objects does not take effect except FT,
because it introduces very frequent data movement which lose performance.
%As a result, our runtime is not able to explore the opportunities exposed
%by partitioning large data objects except in FT.
In FT, this technique contributes to 58\% performance improvement,
while the other three techniques make 42\% contribution by manipulating
small data objects. In general, by this study we learn the importance of combining all techniques to maximize performance improvement for various HPC workloads.

\begin{table}
        \begin{center}
        \caption{Data migration details for HMS with Unimem (NVM has 1/2 DRAM bandwidth).}
        \vspace{-10pt}
        \label{tab:data_mig_details}
        \tiny
        \begin{tabular}{|p{1.2cm}|p{1.3cm}|p{2cm}|p{1.5cm}|p{0.8cm}|}
        \hline
        \textbf{Benchmark}    & \textbf{Times of Migration}  & \textbf{Migrated data size (MB)} &\textbf{Pure runtime cost} & \textbf{\% overlap}                                  \\ \hline \hline
         CG & 3 & 132 & 0.5\% & 66.7\%   \\ \hline
         FT & 4 & 201 & 1.5\% & 75\%   \\ \hline
         BT & 24 & 720 & 1\% & 87.5\%   \\ \hline 
         LU & 3 & 187 & 1\% & 60\%     \\ \hline
         SP & 9 & 348 & 1.5\% & 66.7\% \\ \hline
         MG & 1 & 17 & 2\%	& 100\%				\\ \hline
         Nek5000(eddy) & 102 & 1101 & 3\% & 70.6\% \\ \hline
        \end{tabular}
        \end{center}
        \vspace{-15pt}
\end{table}

To further study the effectiveness of Unimem, we collect some detailed data migration information for HMS with Unimem (NVM has 1/2 DRAM bandwidth). Table~\ref{tab:data_mig_details} shows the results. ``Pure runtime cost'' in the table accounts for the overhead of collecting hardware counters, modeling costs, and synchronization cost between the helper thread and main thread.
``Pure runtime cost'' does not include data movement cost and benefit. ``\% overlap'' in the table shows how many percentage of data movement cost is successfully overlapped with the computation.

From Table~\ref{tab:data_mig_details}, we notice that Unimem has very small runtime overhead (less than 3\% in all cases). Directed by Unimem, the data migration can happen very often (e.g., 102 times in Nek5000 and 24 times in BT), and the migrated data size can be very large (e.g., 1.1GB in Nek5000 and 720MB in BT). However, even with the frequent data migration, Unimem successfully overlaps data migration with computation (70.6\% in Nek5000 and 87.5\% in BT). 
Also, performance benefit of data migration outweighs those non-overlapped data migration, and narrows down the performance gap between NVM and DRAM to 9\% at most (see Figure~\ref{fig:eval_basic_perf_half_dram_bw}).

\begin{figure}[!t]
	\vspace*{5pt}
    \centering
    \includegraphics[width=0.48\textwidth, height=0.13\textheight]{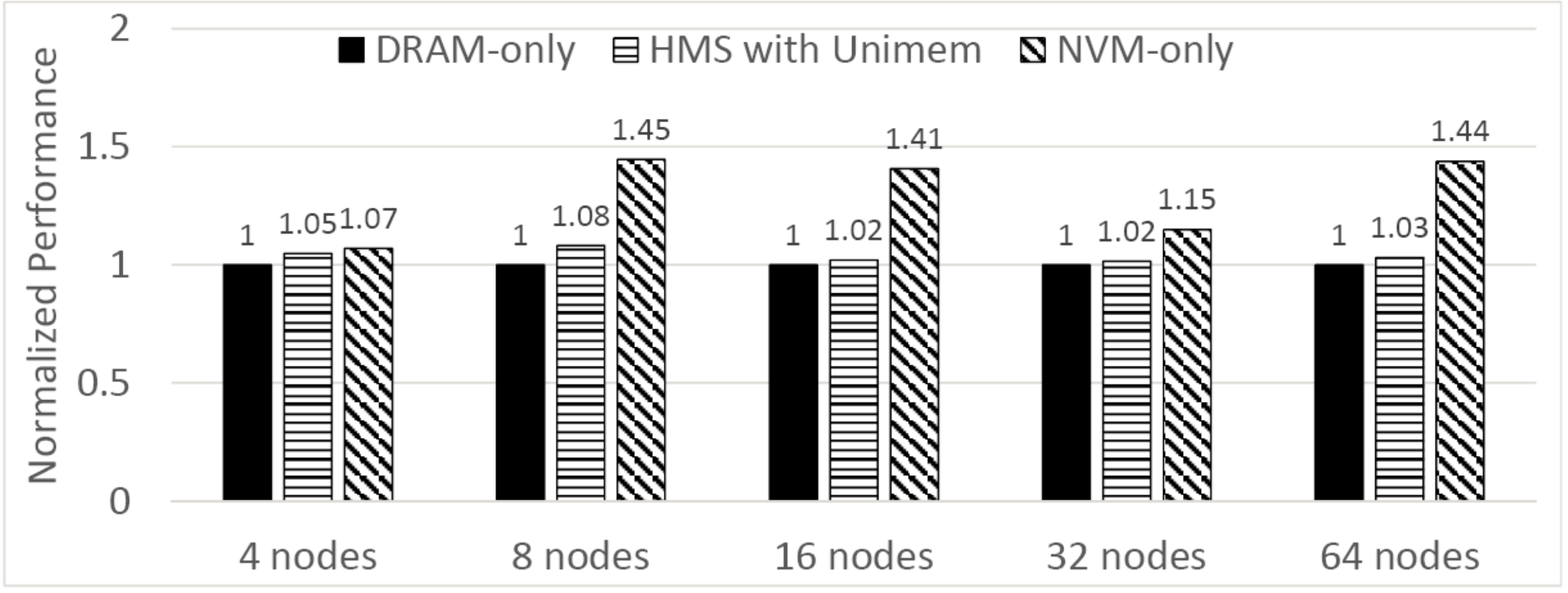}
    \vspace{-20pt}
    \caption{CG strong scaling tests on Edison (LBNL).}
    \label{fig:strong_scaling_cg}
    \vspace{-15pt}
\end{figure}

\textbf{Scalability study.} 
To study how Unimem performs in larger system scales. We did
strong scaling tests on Edison at LBNL. For each test,
we use one MPI task per node and use CLASS D as input problem.
We use 256MB for DRAM and 32GB for NVM. 
Figures~\ref{fig:strong_scaling_cg} %and~\ref{fig:strong_scaling_bt}
shows the results for CG. %and BT.
Performance (execution time) in the figures is normalized to the performance of DRAM-only.
%(strong scaling test): use Edison (LBNL); one MPI task per node; Class D;
%use normal NUMA node; 256MB DRAM

As we change the system scale, the sizes of data objects
change. The numbers of main memory accesses also change because of caching effects: Such changes in main memory accesses impact the sensitivity of data object to memory bandwidth and latency.
Because of the above changes, the runtime system must be adaptive enough
to make a good decision on data placement. 
In general, Unimem does a good job for all cases: the
performance difference between DRAM-only and HMS with Uimem
is no bigger than 7\%. 
%\textbf{Discussion: pending on big data objects.}

\begin{comment}
\begin{figure}[!t]
    \centering
    \includegraphics[width=0.5\textwidth, height=0.15\textheight]{figures/strong_scaling_bt.pdf}
    \caption{BT strong scaling tests on Edison (LBNL).}
    \label{fig:strong_scaling_bt}
    \vspace{-15pt}
\end{figure}
\end{comment}

\begin{figure}[!htbp]
    \centering
    \includegraphics[width=0.48\textwidth, height=0.13\textheight]{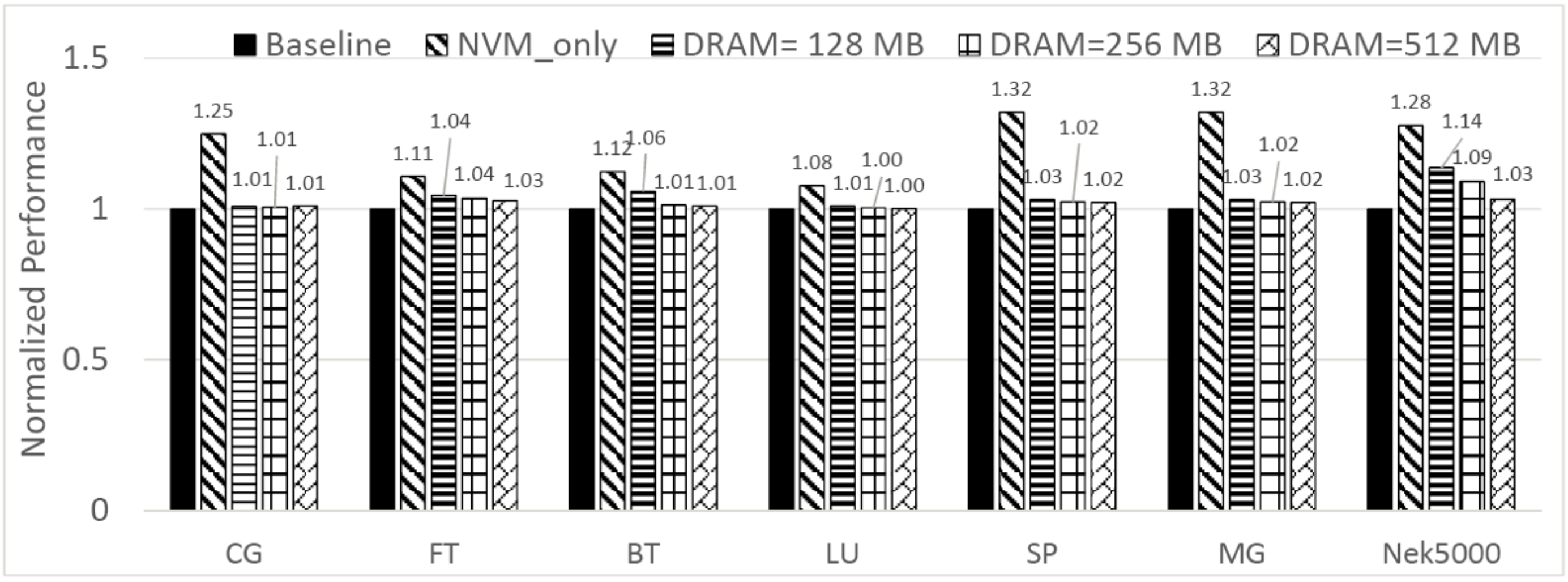}
    \vspace{-20pt}
    \caption{Unimem performance sensitivity to DRAM size in HMS.}
    \label{fig:dram_size_sensitivity}
    \vspace{-15pt}
\end{figure}

\begin{spacing}{0.9}
\textbf{Sensitivity study.}
We use various configurations of DRAM size in HMS and 
test if our runtime can performance well. 
As DRAM size changes, we will have different opportunities 
to place data objects. 
The change of DRAM size will impact the frequency of data movement
and impact whether we should decompose large data objects to improve performance. Figure~\ref{fig:dram_size_sensitivity} shows the results
as we use 128MB, 256MB and 512MB DRAM. In all tests, we use 16GB
NVM configured with 1/2 DRAM bandwidth and CLASS C as input problem. We use Platform A and 4 nodes (1 MPI task per node) to do the tests. In the figure, performance (execution time) is normalized to that of DRAM-only.

In general, Unimem performs well in all cases except one case:
the performance difference between DRAM-only and HMS with Unimem is 
no bigger than 7\% in all cases except MG with 128MB DRAM.
For MG with 128MB DRAM, we have 13\% performance difference between
DRAM-only and HMS with Unimem.
After careful examination, we find that DRAM is not well utilized,
because large data objects cannot be placed in such small DRAM.
We also cannot partition large data objects in MG by using our compiler tool
because of widely employment of memory alias in the benchmark.
But even so, our runtime still narrows performance gap between NVM-only and DRAM-only by 35\%.
\vspace{-5pt}
\end{spacing}

%\vspace{-5pt}

%change the lookahead window size; 
%(1) The impact of phase granularity (local view vs. global view): 
%4 nodes (1 task per node); class C; all 6 NPB benchmarks;
%NVM configuration: (use 1/2 DRAM bandwidth)\\

%(2) The impact of DRAM size: 4 nodes (1 task per node); class C; all 6 NPB %benchmarks; Use 1/2 of DRAM size in Figures 6-7; Use 2X of DRAM size in Figures %6-7; \\

%(3) t1 and t2 thresholds: pending

\section{Related Work}
\label{sec:related_work}
\begin{spacing}{0.9}
Software-managed HMS has been studied in prior work.
Dulloor et. al~\cite{eurosys16:dulloor} introduce a data placement runtime
based on \textit{offline profiling} of application memory access patterns.
Their work targets on enterprise workloads. To decide data placement, they classify memory access patterns into streaming, pointer chasing, and random.
Giardino et. al~\cite{nas16:giardino} rely on OS and application co-scheduling
data placement. In particular, they build APIs that allow programmers
to describe their memory usage characteristics
to OS, through which OS receives and implements
responsive page placement and data migration.
Lin et. al~\cite{asplos16:lin} introduce
a protected OS service for asynchronous memory movement on HMS.
Du et. al~\cite{ismm16:shen} develop a PIN-based \textit{offline profiling} tool
to collect memory traces and provide guidance for placing data on HMS.

Different from the prior efforts, our work requires neither offline profiling as in~\cite{eurosys16:dulloor, ismm16:shen}
nor programmer involvement to identify memory access patterns as in~\cite{nas16:giardino}.
Furthermore, our work does not require the modification of OS, which is different from~\cite{asplos16:lin}.
%Hence, our work provides a better solution for legacy HPC applications and systems.
Our work aims for legacy HPC applications and systems.

Some studies introduce hardware-based data placement solutions for the NVM-based HMS.
Bivens et al.~\cite{hetero_mem_arch} and Qureshi et al.~\cite{qureshi_micro09, ibm_isca09} use DRAM as a set-associative cache logically placed between processor and NVM.
NVM is accessed when DRAM buffer eviction or buffer miss happens. 
Yoon et al.~\cite{row_buffer_pcm_iccd12} place data based on row buffer locality in memory devices.
Wang et al.~\cite{gpu_pcm_pact13} rely on static analysis and advanced memory controller to monitor memory access patterns %at runtime 
to determine data placement on GPU.
Wu et al.~\cite{hpdc16:wu} leverage the knowledge of numerical algorithms to direct data placement. They introduce hardware modifications to support massive data migration and performance optimization. 
Agarwal et al.~\cite{asplos15:agarwal} introduce a bandwidth-aware data placement on GPU, driven by compiler extracted insights and explicit hints from programmers.

A key limitation of the above hardware-based approaches is that they heavily rely on modified hardware to monitor memory access patterns and migrate data.
Some work, such as~\cite{qureshi_micro09, ibm_isca09, gpu_pcm_pact13, row_buffer_pcm_iccd12},
ignores application semantics and triggers data movement based on temporal memory access patterns,
which could cause unnecessary data movement.
Our work avoids any hardware modification,
and explores global optimization on data placement.
%for optimal performance.
\vspace{-15pt}
\end{spacing}

\section{Conclusions}
\begin{spacing}{0.9}
%The emerging NVM provides a solution to address the problem
%of memory scalability and power efficiency for future HPC. However, 
The limitation of NVM imposes a question on 
whether NVM is a feasible solution for HPC workloads. 
In this paper, we quantify the performance gap 
between NVM-based and DRAM-based systems, and demonstrate that
using a carefully designed runtime, it is possible
to significantly reduce the performance gap. 
%We introduce a series of techniques to achieve our design goals, including transparency, lightweight, and non-disruptiveness to hardware and application. 
We hope that our work can lay foundation to embrace NVM for future HPC.
\end{spacing}

%  Use this command to print the description
%
%\printccsdesc

% We no longer use \terms command

%\keywords{ACM proceedings; \LaTeX; text tagging}

%\begin{spacing}{0.9}
%\bibliographystyle{ACM-Reference-Format}
\bibliographystyle{abbrvnat}
\bibliography{li}  % sigproc.bib is the name of the Bibliography in this case
%\end{spacing}

% ACM needs 'a single self-contained file'!
%
%\includepdf[pages=-]{figures/sc17-unimem-artifacts.pdf}

\end{document}